\documentclass[a4paper,11pt]{article}
\pdfoutput=1
\usepackage{jheppub}

\usepackage[utf8x]{inputenc}

\usepackage{subfigure,graphicx,amsmath,amssymb,mathtools,hhline}
\usepackage{color}
\usepackage{float}
\usepackage{multirow}

\usepackage{soul}

\newcommand{\be}{\begin{eqnarray}}
\newcommand{\ee}{\end{eqnarray}}

\newcommand{\ba}{\begin{array}}
\newcommand{\ea}{\end{array}}
\newcommand{\bee}{\begin{equation}\ba{c}}
\newcommand{\eee}{\ea\end{equation}}

\newcommand{\bi}{\begin{itemize}}
\newcommand{\ei}{\end{itemize}}

\newcommand{\lsim}{{\;\raise0.3ex\hbox{$<$\kern-0.75em\raise-1.1ex\hbox{$\sim$}}\;}}
\newcommand{\gsim}{{\;\raise0.3ex\hbox{$>$\kern-0.75em\raise-1.1ex\hbox{$\sim$}}\;}}

\newcommand{\beq}{\begin{equation}}
\newcommand{\eeq}{\end{equation}}
\newcommand{\bea}{\begin{eqnarray}}
\newcommand{\eea}{\end{eqnarray}}
\mathchardef\minus="002D

\usepackage{color}

\toccontinuoustrue

\title{ 
Probing Energetic Light Dark Matter with \\ Multi-Particle Tracks Signatures at DUNE
}
\author[a]{Albert De Roeck,}
\author[b,c]{Doojin Kim,}
\author[d]{Zahra Gh. Moghaddam,}
\author[e]{Jong-Chul Park,}
\author[f]{Seodong Shin,}
\author[g]{and Leigh H. Whitehead}
\affiliation[a]{CERN, Geneva, Switzerland}
\affiliation[b]{Mitchell Institute for Fundamental Physics and Astronomy, Department of Physics and Astronomy, Texas A\&M University, College Station, TX 77843, USA}
\affiliation[c]{Department of Physics, University of Arizona, Tucson, AZ 85721, USA}
\affiliation[d]{Department of Physics, University of Perugia, Perugia, 06123, Italy}
\affiliation[e]{Department of Physics and Institute of Quantum Systems (IQS), Chungnam National University, Daejeon 34134, Republic of Korea}
\affiliation[f]{Department of Physics, Jeonbuk National University, Jeonju, Jeonbuk 54896, Republic of Korea}
\affiliation[g]{Cavendish Laboratory, University of Cambridge, Cambridge, CB3 0HE, United Kingdom}
\emailAdd{albert.de.roeck@cern.ch}
\emailAdd{doojin.kim@tamu.edu} 
\emailAdd{zahra.ghorbani.moghaddam@cern.ch}
\emailAdd{jcpark@cnu.ac.kr} 
\emailAdd{sshin@jbnu.ac.kr}
\emailAdd{leigh.howard.whitehead@cern.ch}

\abstract{
The search for relativistic scattering signals of cosmogenic light dark matter at terrestrial detectors has received increasing attention as an alternative approach to probe dark-sector physics.
Large-volume neutrino experiments are well motivated for searches of dark matter that interacts very weakly with Standard Model particles and/or that exhibits a small incoming flux. 
We perform a dedicated signal sensitivity study for a detector similar to the one proposed by the DUNE Collaboration for cosmogenic dark-matter signals resulting from a non-minimal multi-particle dark-sector scenario.
The liquid argon time projection chamber technology adopted for the DUNE detectors is particularly suited for searching for complicated signatures owing to good measurement resolution and particle identification, as well as $dE/dx$ measurements to recognize merged tracks.
Taking inelastic boosted dark matter as our benchmark scenario that allows for multiple visible particles in the final state, we demonstrate that the DUNE far detectors have a great potential for probing scattering signals induced by relativistic light dark matter.
Detector effects and backgrounds have been estimated and taken into account.
Model-dependent and model-independent expected sensitivity limits for a DUNE-like detector are presented.
\newpage
}

\preprint{
\begin{minipage}{2.0cm}
\small
MI-TH-2011\\
\end{minipage}
}

\begin{document}

\maketitle

\section{Introduction}

The origin of dark matter as observed by astrophysical and cosmological measurements through the gravitational interaction is a strong motivation for  physics beyond the Standard Model (SM).
A plethora of experimental endeavors in search of dark-matter candidates have been made in the last few decades, using direct and indirect detection strategies as well as searches at collider experiments, and mostly focusing on the weakly interacting massive particle (WIMP) paradigm, among other possible candidates. 
Experimental designs and detection schemes often aim for an optimal sensitivity to signals induced by WIMPs. 
No conclusive evidence has been found for dark-matter signals via non-gravitational interactions so far, setting stringent limits over a wide range of the relevant parameter space in dark-matter models. 
This situation presents an opportunity to seriously consider alternative ideas and  methods for searching for dark-matter signals. 
In particular, most of today's dark-matter direct search experiments aim to observe 
\begin{itemize}
\item a {\it nucleus} recoil caused by an {\it elastic} scattering of {\it non-relativistic} dark matter with a {\it weak-scale} mass, 
\end{itemize}
where the absence of an observation may simply mean that we have not yet reached a sufficiently large signal sensitivity.
In this case, increasing the fiducial volume of detectors will allow to scout further the yet non-excluded regions of parameter space.

In this paper, however, we take an alternative approach based on different assumptions as compared to these conventional dark-matter searches.  
More specifically, we will explore a search for   
\begin{itemize}
\item {\it inelastic} scattering processes of {\it boosted} dark matter, i.e., relativistic dark matter, produced in the universe at the {\it present} time, with a {\it non-weak-scale} mass (e.g., MeV to sub-GeV range) in channels with an {\it electron} 
or {\it nucleon} recoil.
\end{itemize}

First of all, we discuss where such a search strategy becomes relevant.
Theoretically, one can envisage the following scenario: while (non-relativistic) cosmological dark matter is still ``thermally'' produced, it is actually secluded from interactions with the SM-sector particles so that it evades detection in direct search experiments.
An example of such a possible scenario is a two-component dark-matter model as proposed in refs.~\cite{Belanger:2011ww,Agashe:2014yua}.
The heavier dark-matter particle (say, $\chi_0$) is assumed to have no direct coupling to SM particles, but instead a lighter dark-matter particle (say, $\chi_1$)  {\it does} directly communicate with SM particles. 
The relic abundance of each component is determined by the ``assisted freeze-out'' mechanism~\cite{Belanger:2011ww} which typically forces $\chi_0$ and $\chi_1$ to be the dominant and negligible relic components, respectively. 
It is then clear why conventional WIMP detectors have not observed (non-relativistic) $\chi_0$ and $\chi_1$ relics: $\chi_0$ comes with large statistics but has suppressed coupling to SM particles while $\chi_1$ comes with a sizable coupling to SM particles but has a negligible amount in the universe.
Note that additional unstable heavier dark-sector particles may exist in such models as well.

The model allows for a sizable interaction between $\chi_0$ and $\chi_1$.
For example, in the annihilation scenario, a pair of $\chi_0$ can annihilate to a pair of $\chi_1$ particles and as a consequence, $\chi_1$ becomes significantly {\it boosted} due to the mass hierarchy between the two particle species.
These $\chi_1$ particles constitute the ``boosted dark matter (BDM)''. 
Hence, relativistic $\chi_1$ scattering processes open up as search channels for dark-matter signals.  
For completeness we note that there are various other ways for creating boosted dark-matter particles, some of which do not require multiple dark-matter particle species; 
for example, semi-annihilating dark matter~\cite{DEramo:2010keq}, fast-moving dark matter~\cite{Huang:2013xfa}, two-component BDM with decaying $\chi_0$~\cite{Bhattacharya:2014yha,Kopp:2015bfa,Heurtier:2019rkz}, solar-capture-enhanced BDM~\cite{Berger:2014sqa,Kong:2014mia}, dynamical dark-matter model~\cite{BDDM}, and cosmic-ray-induced relativistic dark matter~\cite{Bringmann:2018cvk,Ema:2018bih,Dent:2019krz}. 

In exploring many of these models and scenarios, the expected flux of boosted $\chi_1$ (denoted by $\mathcal{F}_1$) near the earth is an important factor to consider: for example, $\mathcal{F}_1 \sim 10^{-6}~{\rm cm}^{-2} {\rm s}^{-1}$ for $\chi_0$ with mass of 10 GeV in the annihilating two-component dark-matter case~\cite{Agashe:2014yua}.
The magnitude of the flux is not large enough for conventional WIMP detectors to have sufficient signal sensitivity unless the mass of $\chi_0$ is significantly smaller than 10 GeV~\cite{Giudice:2017zke}. 
So, large-volume neutrino detectors, of a kiloton (kt) mass scale or more, are typically better suited for the search for relativistic cosmogenic dark-matter signals, and several phenomenological studies have been conducted for neutrino-based facilities including Super-/Hyper-Kamiokande (SK/HK)~\cite{Agashe:2014yua,Berger:2014sqa,Kong:2014mia,Necib:2016aez,Alhazmi:2016qcs,Kim:2016zjx,Kim:2020ipj}, Deep Underground Neutrino Experiment (DUNE)~\cite{Alhazmi:2016qcs,Necib:2016aez,Kim:2016zjx,BDDM,Kim:2019had,Berger:2019ttc,Kim:2020ipj}, IceCube~\cite{Agashe:2014yua,Bhattacharya:2014yha,Kopp:2015bfa,Kong:2014mia,Kim:2020ipj}, prototype detectors for DUNE~\cite{Chatterjee:2018mej, Kim:2018veo}, and detectors in the Short Baseline Neutrino Program~\cite{Kim:2018veo}. 
On top of these efforts in the theory community, the SK Collaboration has reported a first result on the search for BDM elastically interacting with electrons~\cite{Kachulis:2017nci}. 

In the ``minimal'' elastic BDM scattering scenario where the boosted $\chi_1$ manifests itself as target recoil only in a detector, energetic (atmospheric) neutrinos can be a significant source of background as they often leave only a visible target recoil.
To improve the signal sensitivity and reduce background contamination, the data selection is restricted to point-like sources, augmented with directional information, at the cost of signal statistics.
Examples include searches for BDM originating from the sun~\cite{Berger:2014sqa,Kong:2014mia,Alhazmi:2016qcs,Kim:2018veo,Kachulis:2017nci,Berger:2019ttc} and dwarf galaxies~\cite{Necib:2016aez}.

Alternatively, here a different scenario is studied for which signal events  exhibit particular features that backgrounds cannot easily mimic.
In ref.~\cite{Kim:2016zjx}, a search for {\it in}elastic boosted dark matter ($i$BDM) is proposed.
The model allows a boosted $\chi_1$ to scatter off target material and to produce a heavier unstable dark-sector particle (say $\chi_2$), where the mass of $\chi_2$ is larger than the one of $\chi_1$.
By construction, such $\chi_2$ can then decay back to $\chi_1$ and other particles, some of which may be detectable. 
Hence, the expected signatures feature not only the target recoil but additional visible particles in the final state giving more handles to identify signal events. 
A high multiplicity of visible particles in the final state is a natural consequence of many non-minimal dark-sector models. 
While the aforementioned minimal $i$BDM scenario allows for a few additional visible particles, more complex dark-sector scenarios may give rise to a multitude of visible final-state particles via, for example, a cascade decay of a produced heavier dark-sector state.
Experimental signatures with such additional distinctive features are an excellent motivation for searches using high quality and performance detectors, equipped with good energy/angular/position resolution and particle identification.
The first $i$BDM signal search was performed by a dark matter direct detection experiment, COSINE-100~\cite{Ha:2018obm}.

We investigate in this paper the detection potential of multi-particle signals in the DUNE far detectors~\cite{Abi:2020wmh,Abi:2020evt,Abi:2020loh,Abi:2018rgm}, taking inelastic boosted dark matter as the benchmark scenario.
A preliminary study was performed in ref.~\cite{Kim:2016zjx} for a zero background assumption.
Similarly, ref.~\cite{Chatterjee:2018mej} discussed the $i$BDM sensitivity for a search using the DUNE prototype (ProtoDUNE) detectors, including a careful estimate of the potential background events.
Note that these ProtoDUNE detectors are located on surface and hence are exposed to a vast cosmic-origin background, whereas the DUNE far detectors will be installed deep underground strongly reducing this background.
Nevertheless, we carefully examine possibilities that could give rise to signal-like background events, and at the same time identify selection criteria to achieve a vanishing background with good signal efficiency.

For definiteness we adopt a dark-photon scenario to take care of the interactions between (boosted) $\chi_1$ and SM particles as our benchmark model.
Since the model, in principle, does not impose any particular preference for the boosted $\chi_1$ to scatter off electrons or protons, we will study both electron and proton scattering channels, which can be complementary especially at the earlier stages of experiments~\cite{Kim:2020ipj}.
We will discuss how dark-photon model parameters can be constrained by the DUNE experiment, in the context of $i$BDM searches. 

This paper is organized as follows. 
We begin in section~\ref{sec:model} by defining and discussing the $i$BDM signal, i.e., boosted dark matter production and its interactions with SM particles, followed by explaining the expected experimental signatures in section~\ref{sec:signatures}.
We then briefly summarize key characteristics of the DUNE far detectors and discuss potential background events to the $i$BDM signal search in section~\ref{sec:detector}.
Section~\ref{sec:select} deals with selection criteria applicable to DUNE or DUNE-type detectors.
We discuss the advantages from the capability of measuring the $dE/dx$ of charged particles, in particular, for merged multi-track events.
Based on the signal selection, we present phenomenological studies for both the electron and the proton scattering channels in section~\ref{sec:results}. 
We first discuss model-dependent sensitivity reaches, the coverage in dark-photon parameter space and the experimental reach of the velocity-averaged annihilation cross section of $\chi_0$ as a function of the $\chi_0$ mass, in section~\ref{sec:modeldep}.
In addition, section~\ref{sec:modelind} is reserved for possible model-independent sensitivity reaches expected for the DUNE far detectors.
Finally, conclusions are presented in section~\ref{sec:conclusion}.

\section{The Model}

Here we review our benchmark dark-matter model for the production of boosted dark matter in the universe today, and its coupling to SM particles. 
The model dependence of the latter defines the expected experimental signatures.

\subsection{Benchmark scenario \label{sec:model}}

In our model, the dark sector contains (at least) two different dark-matter particles that are stable as a result of protection by unbroken separate symmetries such as $Z_2 \otimes Z_2'$ and ${\rm U}(1)' \otimes {\rm U}(1)''$ in, e.g., the model as described in ref.~\cite{Belanger:2011ww}.
The model further assumes that one of the two dark-matter species (typically the heavier one, $\chi_0$) does {\it not} directly interact with the SM particles, whereas the other 
one ($\chi_1$) {\it does} interact with the SM particles. On the other hand, interactions between $\chi_0$ and $\chi_1$ are allowed; for example, the $\chi_0$ may pair-annihilate to a $\chi_1$ pair. The indirect coupling of $\chi_0$ to SM particles (through $\chi_1$) is typically loop-suppressed. 

The  $\chi_0$ relic abundance relative to the $\chi_1$ one is assumed to be governed by the ``assisted freeze-out'' mechanism~\cite{Belanger:2011ww}.
Due to the model setup, $\chi_0$ is not in direct contact with the thermal bath, but has thermalized through the ``assistance'' of $\chi_1$. 
In typical cases, $\chi_0$ froze out earlier, and hence became the dominant relic playing the role of cosmological dark matter, while $\chi_1$ froze out later, and ended up constituting a negligible amount of the overall dark-matter abundance. 
As mentioned earlier, standard dark-matter direct detection experiments  
are typically not sensitive yet to detect either $\chi_0$ or $\chi_1$ due to suppressed coupling to SM and small relic contribution, respectively.
However, $\chi_1$ can be boosted by pair-annihilation of the (non-relativistic) $\chi_0$ in the universe today and therefore searching for relativistic scattering signatures induced by boosted $\chi_1$ is of interest  to pursue~\cite{Agashe:2014yua}.
 
First, we estimate the expected flux of $\chi_1$ near the earth:
\bea
\mathcal{F}_1&=& \frac{1}{2}\cdot \frac{1}{4\pi}\int d\Omega \int_{\rm l.o.s.} ds \langle \sigma v \rangle _{\chi_0\bar{\chi}_0 \rightarrow \chi_1\bar{\chi}_1} \left(\frac{\rho(s,\theta)}{m_0}\right)^2 \nonumber \\
&=& 1.6\times 10^{-6}~{\rm cm}^{-2}{\rm s}^{-1}\times \left(\frac{\langle \sigma v \rangle _{\chi_0\bar{\chi}_0 \rightarrow \chi_1\bar{\chi}_1}}{5\times 10^{-26}~{\rm cm}^3\,{\rm s}^{-1}} \right)\times \left( \frac{10~{\rm GeV}}{m_0}\right)^2\,, \label{eq:fluxform}
\eea
where $m_0$ denotes the mass of $\chi_0$, $\rho$ describes the $\chi_0$ density profile as a function of the line-of-sight (l.o.s.) $s$ and solid angle $\Omega$, and $\langle \sigma v \rangle _{\chi_0\bar{\chi}_0 \rightarrow \chi_1\bar{\chi}_1}$ stands for the velocity-averaged annihilation cross section for the $\chi_0\bar{\chi}_0 \rightarrow \chi_1\bar{\chi}_1$ process in the universe
today.
We assume here that $\chi_0$ and its antiparticle $\bar{\chi}_0$ are distinguishable, thus the pre-factor 1/2 can be dropped for the indistinguishable case. 
In order to calculate the numerical value for $\mathcal{F}_1$  we apply the  Navarro-Frenk-White (NFW) dark-matter halo profile~\cite{Navarro:1995iw,Navarro:1996gj} for $\rho(s,\theta)$  with local dark-matter density $\rho_\odot = 0.3~{\rm GeV}\cdot{\rm cm}^{-3}$ near the sun which is 8.33~kpc away from the galactic center, the scale density $\rho_s=0.184~{\rm GeV}\cdot{\rm cm}^{-3}$, scale radius $r_s=24.42$~kpc, and slope parameter $\gamma=1$. We then take 10 GeV and $5\times 10^{-26}~{\rm cm}^3\,{\rm s}^{-1}$ as reference values for $m_0$ and $\langle \sigma v \rangle _{\chi_0\bar{\chi}_0 \rightarrow \chi_1\bar{\chi}_1}$. 
The chosen value for the present-day velocity-averaged annihilation cross section agrees with the observed dark-matter density, which is valid for BDM scenarios in which the dominant relic abundance is set by the $s$-wave annihilation process $\chi_0\bar{\chi}_0\rightarrow \chi_1\bar{\chi}_1$.
One may impose an angular cut depending on the scope of the analysis, but here we assume that the data is collected over the {\it whole sky} throughout this paper.\footnote{The majority of boosted $\chi_1$ is expected to come from the galactic center.} 

Next, for the interactions between $\chi_1$ and the SM-sector particles, we use a vector portal scenario where a massive dark-sector photon (denoted as $X$) is the new gauge boson of the dark gauge symmetry U(1)$_X$.
With fermionic dark-matter particles for illustration, the relevant interacting Lagrangian can be written as: 
\bea
-\mathcal{L}_{\rm int} \supset \frac{\epsilon}{2}X_{\mu\nu}F^{\mu\nu}+g_{11}\bar{\chi}_1\gamma^{\mu}\chi_1 X_\mu +  g_{12}\bar{\chi}_2\gamma^{\mu}\chi_1 X_\mu +h.c., \label{eq:laglangian}
\eea
where $\epsilon$ is the parameter of the  kinetic mixing between U(1)$_X$ and U(1)$_{\rm SM}$, and $X_{\mu\nu}$ ($F_{\mu\nu}$) is the field strength tensor for the dark-sector (SM-sector) photon.
We here introduce another symbol $\chi_2$ to represent a heavier unstable dark-sector state, that is, the mass of $\chi_2$, $m_2$, is larger than that of $\chi_1$, $m_1$.
The second and the third operators are responsible for elastic scattering and inelastic scattering of $\chi_1$, respectively, and $g_{11}$ and $g_{12}$ parameterize the associated coupling strengths.
As stated before, the main focus in this paper is the search for inelastic BDM, but we include the second term as well for completeness.

A few comments are in order.
First, there are six model parameters relevant to the $i$BDM search: $\epsilon$, $g_{12}$, $m_0$, $m_1$, $m_2$, and $m_X$, with $m_X$ being the mass of dark photon.
So, it is tedious to interpret data and present results in a verbose way.
Therefore, we will fix some of the parameters motivating our choice in section~\ref{sec:results}.
Second, it is possible to build $i$BDM-dominating models where $g_{11}$ is either highly suppressed or even vanishing compared to $g_{12}$
 (see e.g., Appendix A of ref.~\cite{Giudice:2017zke} for a more concise and systematic discussion).
Finally, in the annihilating two-component dark-matter scenario under consideration here, $m_0$ is the same as the energy of the boosted $\chi_1$, $E_1$.
We shall use $m_0$ and $E_1$ interchangeably throughout this paper.

\subsection{Experimental signatures \label{sec:signatures}}

\begin{figure}[t]
\centering
\includegraphics[scale=0.7]{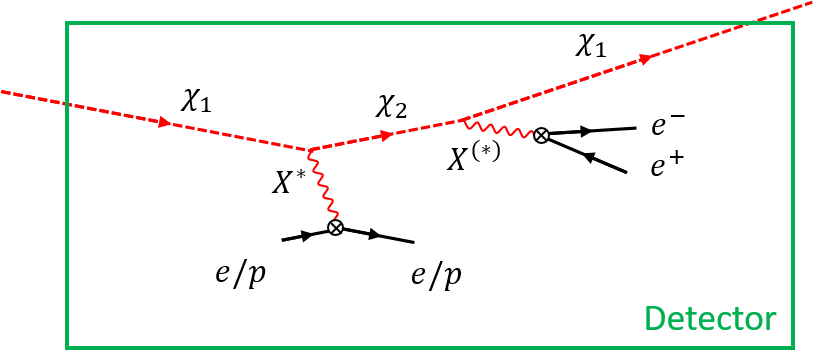}
\caption{\label{fig:signature} A schematic presentation
of the experimental signatures for this study. The primary scattering arises via the exchange of a virtual dark photon, resulting in a proton or electron recoil, and the produced $\chi_2$ subsequently decays back to a $\chi_1$ and a $e^+e^-$ pair through either an on-shell or off-shell dark-photon exchange, depending on the underlying mass spectrum. The interaction point of the electron-positron pair can be visibly displaced from that of the target recoil.  
}
\end{figure} 

The dark-photon in eq.~\eqref{eq:laglangian} can interact with two targets in a detector medium: electrons and protons.
We approximate the proton to be a free nucleon in the energy region that we study. 
Collisions on protons can lead to deep inelastic scattering (DIS) processes if the associated momentum transfer is sufficiently sizable. 
However, in most of the motivated parameter search space (e.g., MeV to sub-GeV range in $m_X$) the contribution from DIS is subdominant~\cite{Kim:2020ipj}.
Hence, in this study,  the proton is taken as a composite particle with a nontrivial internal structure instead of constituent partons.  
The dark-matter interaction process that we consider is given by the initial scattering
\bea
\chi_1 + e^-/p \rightarrow e^-/p +\chi_2,\label{eq:primscatter}
\eea
followed by the subsequent decay of $\chi_2$
\bea\chi_2 \rightarrow \chi_1 X^{(*)} \rightarrow \chi_1 e^+e^-,\label{eq:chi2decay}
\eea
and is depicted in figure~\ref{fig:signature}, 
i.e., {\it three} charged particles will emerge in the fiducial volume of the  detector.\footnote{Related searches can be conducted at fixed target experiments with active production of relativistic  $\chi_1$~\cite{deNiverville:2011it,Izaguirre:2014dua,Kim:2016zjx}.}
The primary scattering process shown in eq.~\eqref{eq:primscatter} arises via exchange of a virtual dark photon. 
In the secondary process given in eq.~\eqref{eq:chi2decay}, $\chi_2$ decays back to $\chi_1$ and an electron-positron pair. 
This happens via either on-shell or off-shell dark-photon exchange, depending on the underlying mass spectrum and ``$X^{(*)}$'' 
represents these two possibilities. 
If $m_X$ is larger than $m_2-m_1$, $\chi_2$ decay via a three-body decay process (i.e., off-shell $X$ exchange), whereas in the case of $m_2>m_X+m_1$, $\chi_2$ decays to a $\chi_1$ and an on-shell $X$.\footnote{To ensure a visible decay of $X$ in the latter case, $m_X$ should be smaller than $2m_1$.}
Note that the $e^+e^-$ pair could be replaced by a generic SM fermion pair $f\bar{f}$ whenever kinematically allowed, but this paper concentrates on the $e^+e^-$ final state for definiteness.   

\begin{figure}[t]
\centering
\includegraphics[width=7.3cm]{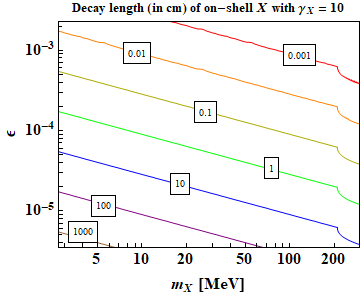} \hspace{0.2cm}
\includegraphics[width=7.3cm]{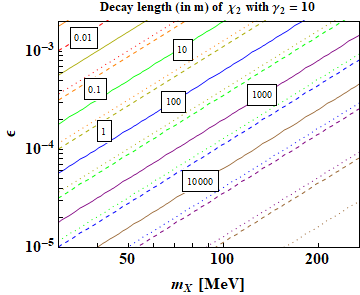}
\caption{\label{fig:contour} Left: Decay length of $X$ in centimeters in the $m_X-\epsilon$ plane for the case where $\chi_2$ decays to $\chi_1$ and an on-shell $X$.
The Lorentz boost factor of $X$ is set to be 10 for illustration.
Right: Decay lengths of $\chi_2$ in meters in the $m_X-\epsilon$ plane for the case where $\chi_2$ decays to $\chi_1$ and an electron-positron pair via an off-shell $X$. The Lorentz boost factor of $\chi_2$ is again set to be 10 for illustration.
Three different values of mass difference $(m_2-m_1)$ are shown: 10 MeV (solid lines), 20 MeV (dashed lines), and 30 MeV (dotted lines). }
\end{figure}

An intriguing possibility for the signal events is that the $e^+e^-$ pair from the decay in eq.~\eqref{eq:chi2decay} can be significantly displaced from the primary target recoil vertex.
For the case in which the $\chi_2$ decays to a $\chi_1$ and an on-shell $X$ via a two-body process, it is usually a prompt decay  unless the mass spectrum is extremely degenerate or $g_{12}$ is very
small. 
So, in order to have a displaced vertex the on-shell $X$ would itself have to be long-lived.
The decay width $\Gamma_X$ is expressed as
\bea
\Gamma_X = \frac{\epsilon^2\alpha\, m_X}{3}\left(1+\frac{m_e^2}{m_X^2}\right)\sqrt{1-\frac{4m_e^2}{m_X^2}}\,,
\eea
which can be translated to the laboratory-frame mean decay length $\ell_{X,{\rm lab}}$
\bea
\ell_{X,{\rm lab}} \sim 40\, {\rm cm}\cdot \left(\frac{10^{-5}}{\epsilon}\right)^2\left(\frac{20\,{\rm MeV}}{m_X}\right)\frac{\gamma_X}{10}. \label{eq:lX}
\eea
Here $\alpha$ is the usual electromagnetic fine structure constant and  $\gamma_X$ is the Lorentz boost factor of $X$.
We show the contour plot for eq.~\eqref{eq:lX} in the plane of $m_X$ and $\epsilon$ in the left panel of figure~\ref{fig:contour}.
The contour values are in centimeters.
This relation shows that small $m_X$ and $\epsilon$ values favor sizable decay lengths.
However, small $\epsilon$ values reduce the primary scattering cross section and small $m_X$ values are disfavored by the current limits. 
Therefore, ``displaced'' $i$BDM events by a long-lived $X$ are phenomenologically less favored. 

On the other hand, $\chi_2$ can be long-lived if it undergoes a three-body decay. 
The corresponding decay width $\Gamma_2$ is given by
\bea
\Gamma_2 \approx \frac{\epsilon^2\alpha g_{12}^2}{15\pi^2m_X^4}(m_2-m_1)^5\,,
\eea
where we assumed the hierarchy of $m_e \ll m_2-m_1 \ll m_2 \ll m_X$. 
We provide the exact formula in the appendix~\ref{sec:app}.
Mapping this expression to the laboratory-frame mean decay length of $\chi_2$ (denoted as $\ell_{2,{\rm lab}}$), we have
\bea
\ell_{2,{\rm lab}} \sim 62\,{\rm cm}\cdot \left(\frac{10^{-3}}{\epsilon}\right)^2 \left(\frac{1}{g_{12}}\right)^2\left(\frac{m_X}{100\,{\rm MeV}}\right)^4\left(\frac{20\,{\rm MeV}}{m_2-m_1}\right)^5\frac{\gamma_2}{10}\,,\label{eq:l2}
\eea
where $\gamma_2$ denotes the Lorentz boost factor of $\chi_2$. 
This expression demonstrates that there exists an interesting range of parameter values for which $i$BDM events have significantly displaced interaction vertices within the 
(fiducial) detector volume. 
In particular, it is encouraging that a small value of $\epsilon$ is {\it not} required, unlike the case for the long-lived $X$, which implies that the ``displaced'' $i$BDM signal in this case can have a sizable scattering cross section.
The contour plots for eq.~\eqref{eq:l2} in the plane of $m_X$ and $\epsilon$ are shown in the right panel of figure~\ref{fig:contour}, with the contour values given in meters. Three different values of mass difference $(m_2-m_1)$ are shown for illustration: 10 MeV (solid lines), 20 MeV (dashed lines), and 30 MeV (dotted lines).

\subsection{Kinematic features \label{sec:kin}}

In this subsection, we discuss the maximum mass reach of $\chi_2$ and two important experimental observables of signal events: energy spectra and angular correlations of final-state (visible) particles. 
The energy spectrum of the final state particles is an important characteristic 
of the process as it drives the probability that the energy deposits will pass the detection thresholds.
On the other hand, the angular correlation observable is connected to the angular resolution since unresolvable angular separation will potentially lead to signal events with merged particle tracks.  
To study the distributions of particle energies and angular separation, we developed our own Monte Carlo simulation code using the primary-scattering matrix element and fully implementing the secondary-decay matrix element, as will be discussed below.  

As for any accelerator experiment, the maximum value of $m_2$ is $\sqrt{s}-m_T$ where $T$ stands for target particle, i.e., $T=e$ or $p$, and where $s$ is the center-of-mass energy given by $s=m_T^2+2E_1m_T+m_1^2$.
In other words, we have
\begin{equation}
    m_2 \leq \sqrt{m_T^2+2E_1 m_T+m_1^2}-m_T\,.
\end{equation}
Two extreme cases are considered.
If $m_1$ is much greater than $m_T$ -- which is usually the case for electron scattering -- along with a
sizable Lorentz boost factor for $\chi_1$, the above relation is approximately
\begin{equation}
    m_2 \leq m_1+(\gamma_1-1)m_T\,. \label{eq:m2elec}
\end{equation}
The opposite limiting case where $m_1$ is much smaller than $m_T$ -- which is usually the case for proton scattering -- results in
\begin{equation}
    m_2 \leq \gamma_1 m_1\,. \label{eq:m2prot}
\end{equation}
Equations \eqref{eq:m2elec} and \eqref{eq:m2prot} imply that the proton scattering channel is more effective than the electron scattering channel in probing dark-sector states that are much heavier than the incoming dark matter $\chi_1$. 

The differential spectrum of recoiling particles $d\sigma/dE_T$ are given by~\cite{Kim:2016zjx}
\begin{align}
    \frac{d\sigma}{dE_T} = \frac{4\alpha \epsilon^2 g_{12}^2 m_T^2}{ \lambda(s,m_T^2, m_1^2)\left\{2m_T(E_2-E_1)^2-m_X^2\right\}^2} &\left[ \mathcal{M}_0(F_1+\kappa F_2)^2+ \mathcal{M}_1\left\{ -\kappa F_2(F_1+\kappa F_2) \right. \right. \nonumber \\
    & \left. \left. + (\kappa F_2)^2 \frac{E_1-E_2+2m_T}{4m_T}\right\} \right]\,, \label{eq:diffscat}
\end{align}
where $\lambda$ is the usual kinematic triangular function defined as $\lambda(x,y,z)=(x-y-z)^2-4yz$ and where $E_1$ and $E_2$ are the incoming $\chi_1$ energy and the outgoing $\chi_2$ energy, respectively, measured in the laboratory frame. 
The $\chi_2$ energy $E_2$ can be converted to the recoil energy $E_T$ by the relation $E_2=E_1+m_T-E_T$. 
Here $\mathcal{M}_0$ and $\mathcal{M}_1$ are defined as follows:
\begin{eqnarray}
    \mathcal{M}_0&=&\left[ m_T(E_1^2+E_2^2)-\frac{(m_2-m_1)^2}{2}(E_2-E_1+m_T)+m_T^2(E_2-E_1)+m_1^2E_2-m_2^2E_1\right], \nonumber \\
    && \\
    \mathcal{M}_1&=& m_T \left[ \left(E_1+E_2-\frac{m_2^2-m_1^2}{2m_T} \right)^2 +(E_1-E_2+2m_T)\left\{E_2-E_1-\frac{(m_2-m_1)^2}{2m_T} \right\} \right], \nonumber \\
    && \label{eq:M1form}
\end{eqnarray}
where $F_1$ and $F_2$ represent form factors. 
For the electron target, we take $F_1=1$ and $F_2=0$, whereas for the proton target, we adopt  values presented in ref.~\cite{Qattan:2004ht} together with the proton anomalous magnetic moment $\kappa =1.79$
(see Appendix of ref.~\cite{Kim:2020ipj} for more detailed expressions).
Note that one can easily obtain the differential spectrum for the elastic scattering, $\chi_1 +e^-/p \rightarrow \chi_1 +e^-/p$,
from eqs.~\eqref{eq:diffscat} through \eqref{eq:M1form} in the limit of $m_2 \to m_1$ and $g_{12} \to g_{11}$ where $E_2$ is interpreted as the energy of the outgoing $\chi_1$.

\begin{table}[t]
    \centering
    \begin{tabular}{c|c c c c}
    \hline \hline
         & $E_1$ [MeV] & $m_1$ [MeV] & $m_2$ [MeV] & $m_X$ [MeV] \\
    \hline
    REF1 & 2000 & 50 & 60 & 50 \\
    REF2 &  400 &  5 & 15 & 50 \\
    \hline \hline 
    \end{tabular}
    \caption{Mass spectra for two reference points, REF1 and REF2.
    In the annihilating two-component dark-matter scenario, $E_1$ can be identified with $m_0$.
    These are the baseline parameter choices for this study. Depending on the analyses, one or two mass parameters in each reference point will be varied.}
    \label{tab:refpoints}
\end{table}

\begin{figure}[t]
\centering
\includegraphics[width=7.3cm]{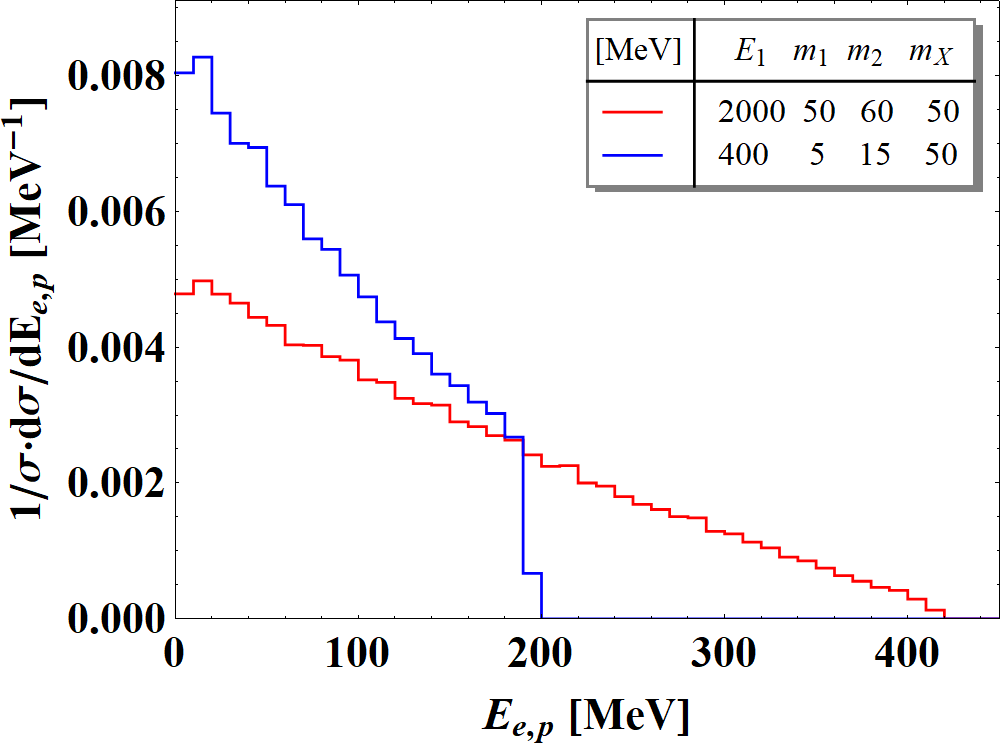} \hspace{0.2cm}
\includegraphics[width=7.3cm]{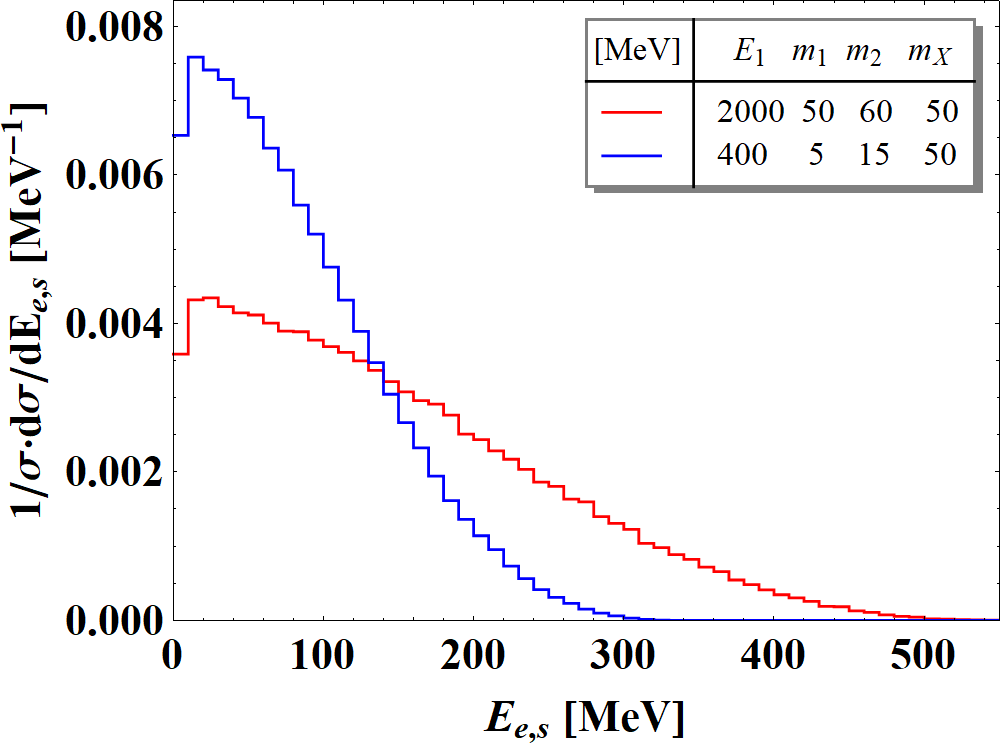} \\ \vspace{0.2cm}
\includegraphics[width=7.3cm]{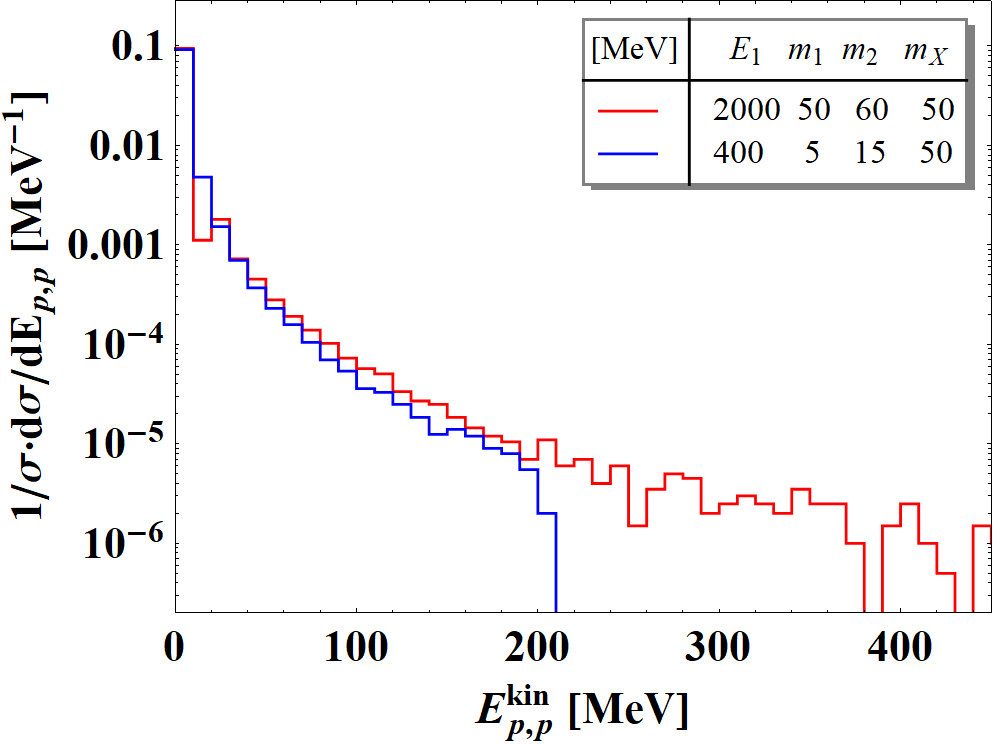} \hspace{0.2cm}
\includegraphics[width=7.3cm]{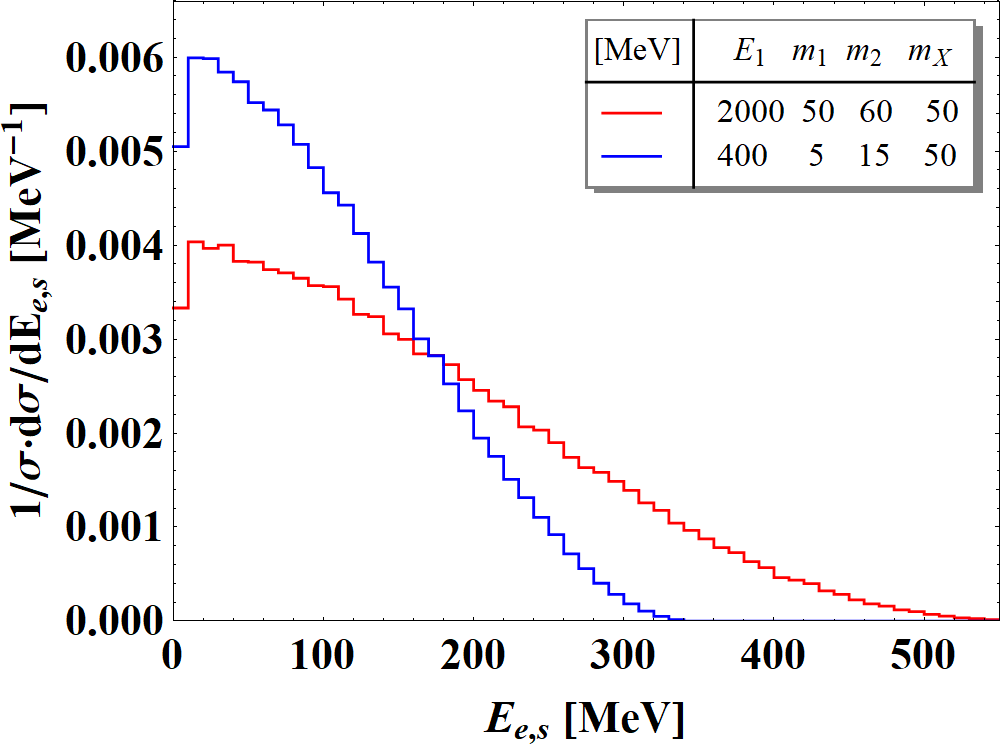}
\caption{\label{fig:espec} Top: Recoil electron energy spectrum (left) and secondary electron energy spectrum (right) for the case of scattering off electrons.
The reference parameter choices, REF1 (red) and REF2 (blue), are shown in the legends.
Bottom: Recoil proton kinetic energy spectrum (left) and secondary electron energy spectrum (right) for the case of scattering off protons.}
\end{figure}

A simple kinematic consideration suggests that the maximally (minimally) allowed recoil energy $E_T^+$ ($E_T^-$) be
\begin{equation}
    E_T^\pm = \frac{(s+m_T^2-m_2^2)(E_1+m_T)\pm \lambda^{1/2}(s,m_T^2,m_2)\sqrt{E_1^2-m_1^2}}{2s}\,. \label{eq:eTrange}
\end{equation}
The left panels of figure~\ref{fig:espec} show the shapes of recoil energy spectra for signal events, generated according to eq.~\eqref{eq:diffscat}.
The top panel and the bottom panel show respectively  the recoil electron energy\footnote{Electrons in our study are energetic enough to interchangeably use energy and kinetic energy.} spectra on a linear scale and the recoil proton kinetic energy spectra on a  logarithmic scale, for two reference mass points, REF1 and REF2, as detailed in the figure legends and in table~\ref{tab:refpoints}.
The main difference between the electron recoil and proton recoil is that for the proton case the energy is not efficiently transferred to the proton target unless $m_1$ is comparable to $m_p$.
Therefore, the typical kinetic energy of the recoiling protons is small, requiring a small detector energy threshold for protons to be detectable.
This is clearly different from the top-left panel of figure~\ref{fig:espec}:
the recoiling protons are more peaked toward smaller values. 

For the energy of visible decay products from the secondary decay, we show the expected energy spectra for the electron scattering case and the proton scattering case in the top-right panel and the bottom-right panel of figure~\ref{fig:espec}, assuming the same benchmark reference points as before. 
For definiteness, the visible decay products are taken to be electrons and positrons. 
These are energetic enough to exceed the energy threshold as will be shown below (table~\ref{tab:detectors} in section \ref{sec:detector}), so these have a high probability to be visible in the detector unless the decay is significantly delayed and is outside the detector acceptance.

\begin{figure}[t]
\centering
\includegraphics[width=7.3cm]{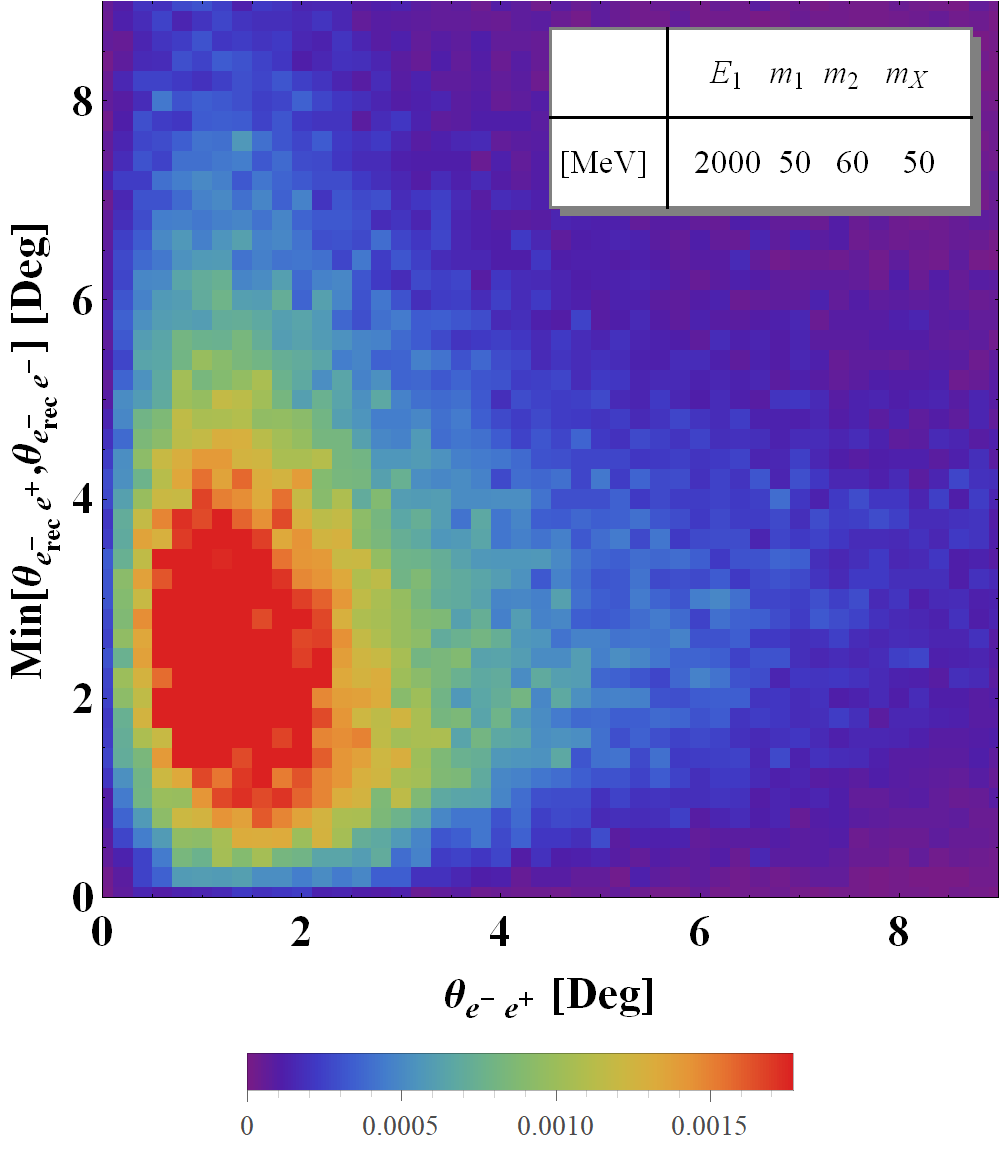} \hspace{0.2cm}
\includegraphics[width=7.3cm]{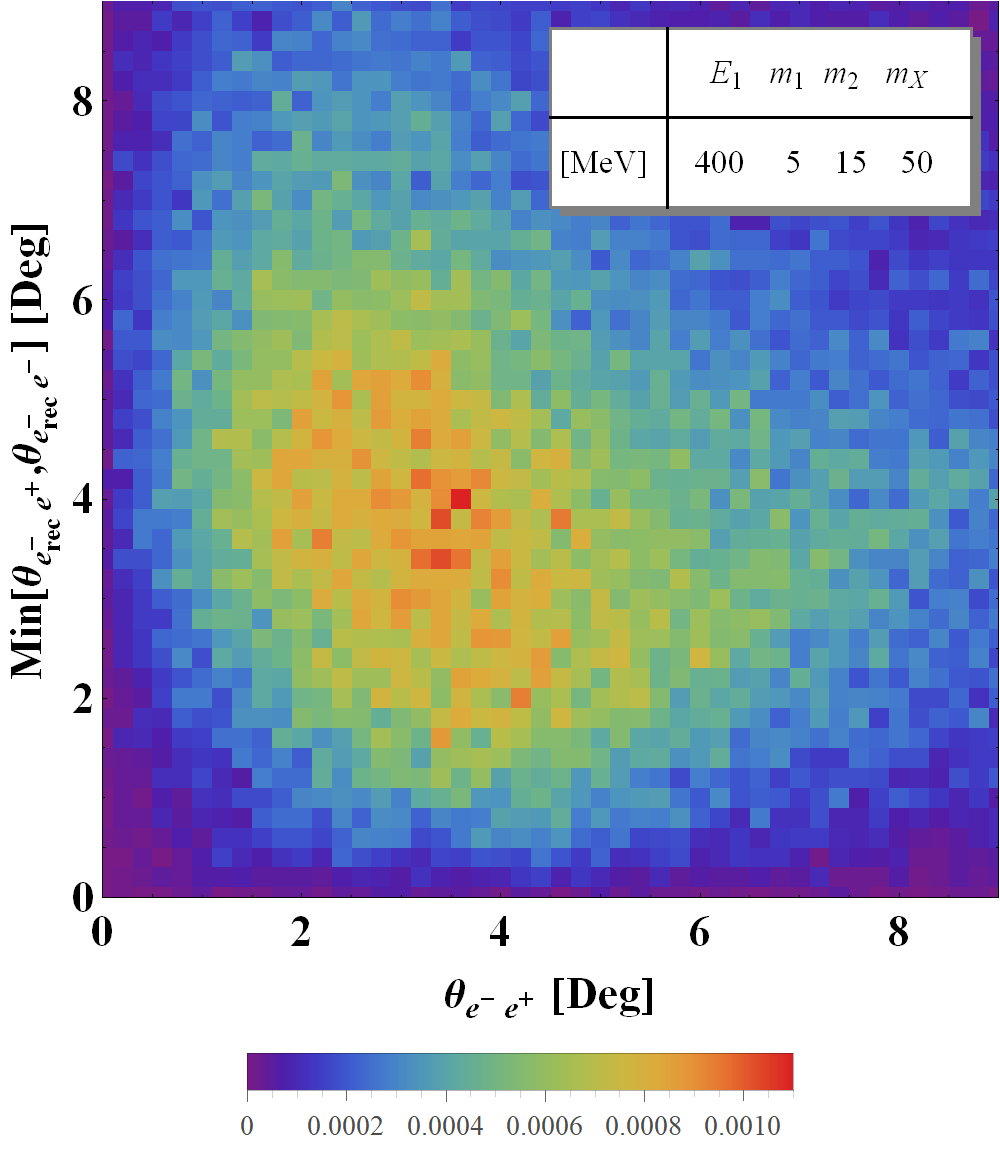} \\ \vspace{0.2cm}
\includegraphics[width=7.3cm]{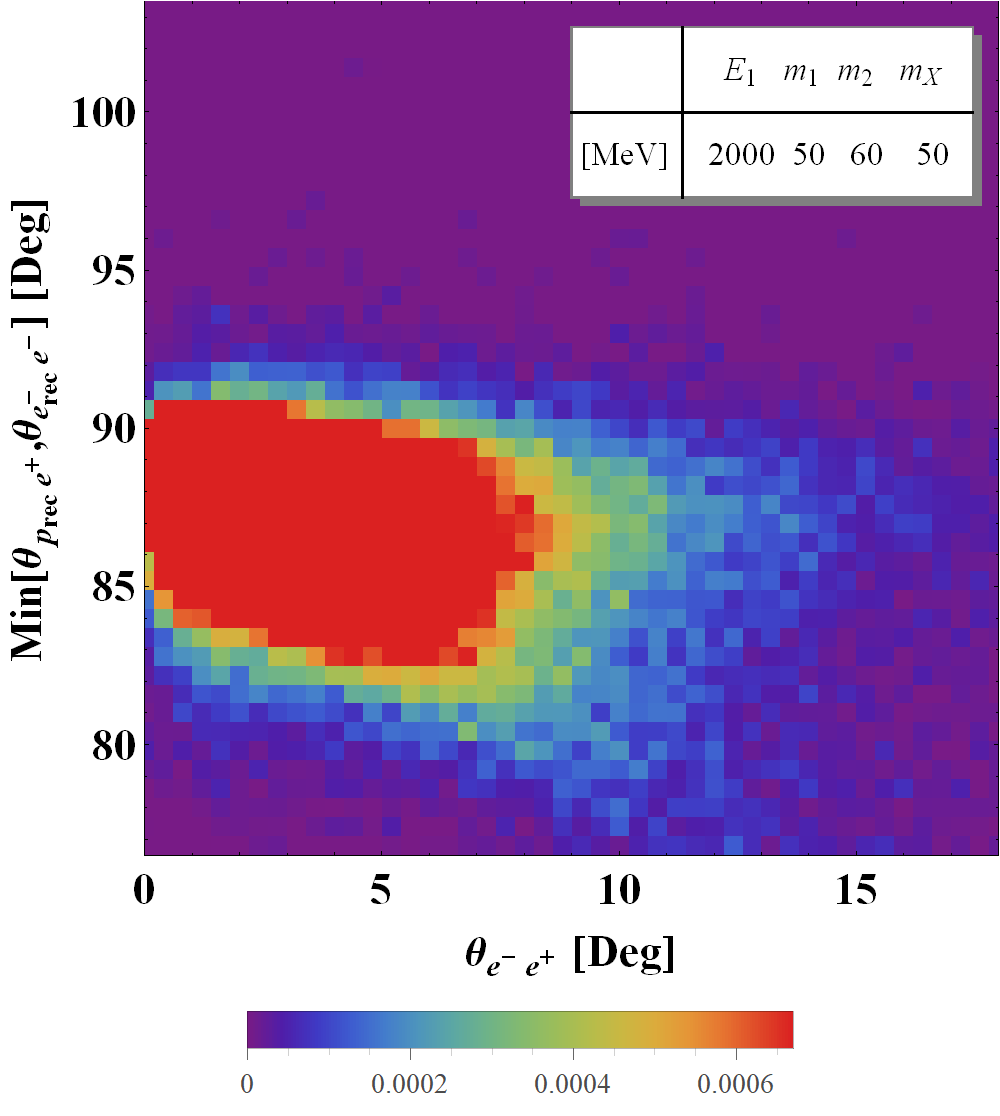} \hspace{0.2cm}
\includegraphics[width=7.3cm]{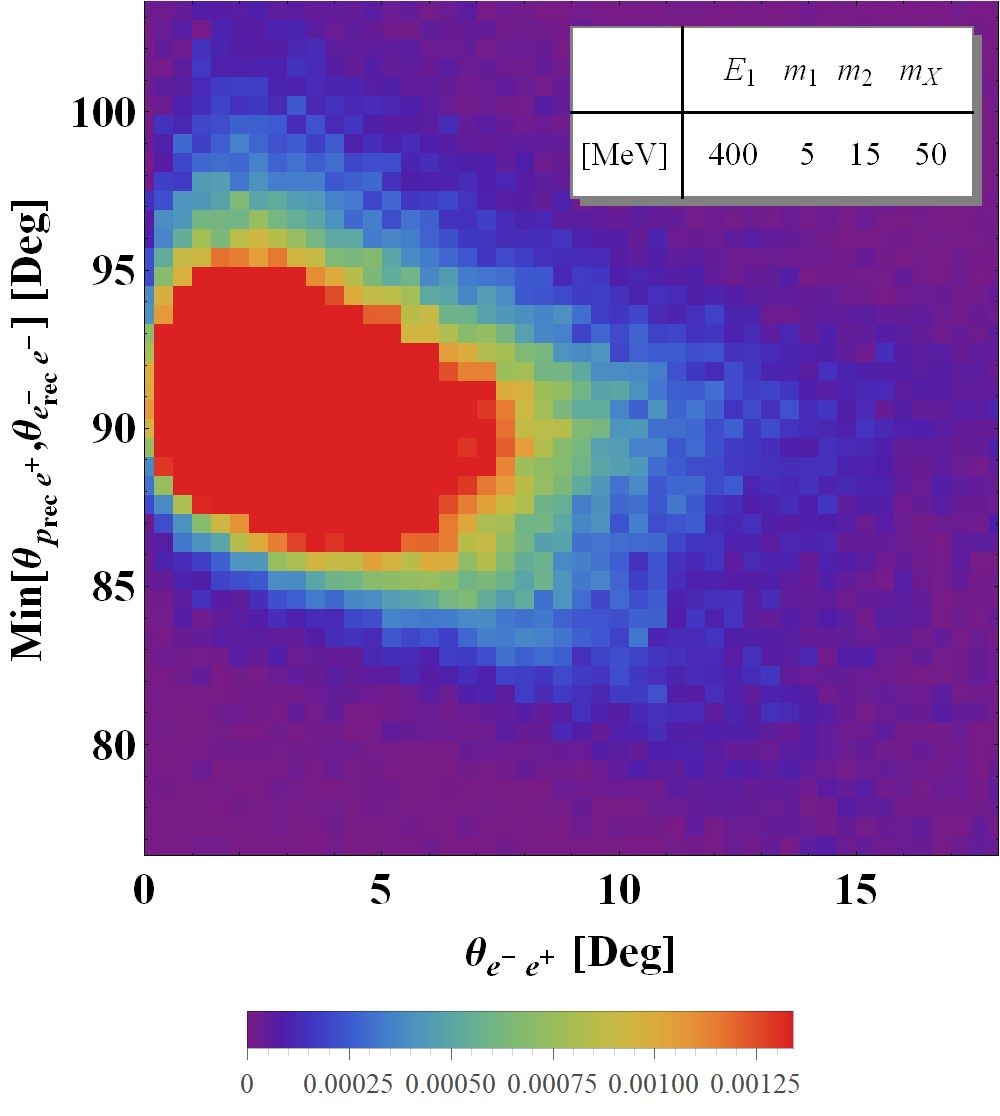}
\caption{\label{fig:angcorr} Scatter plots of the plane of ${\rm Min}\left[\theta_{e_{\rm rec}^-e^-},\theta_{e_{\rm rec}^-e^+}\right]$ and $\theta_{e^-e^+}$ in the electron scattering channel (top panels) and of the plane of ${\rm Min}\left[\theta_{p_{\rm rec}e^-},\theta_{p_{\rm rec}e^+}\right]$ and $\theta_{e^-e^+}$ in the proton scattering channel (bottom panels) with REF1 (left panels) and REF2 (right panels).
All the angles are in degrees and the color code indicates unit-normalized relative event densities.
See the text for variable definitions.}
\end{figure}

As stated earlier, identification of the secondary decay products is crucial for the signature to differentiate from potential background events, and thus for the signal sensitivity. 
As a result of the large boost of the incident $\chi_1$, the visible particles in the final state are often collimated. 
Reliable signal event tagging depends on to what extent we can identify these collimated particles within the angular resolution of the detector.
Denoting the angular separation between $\chi_2$ and the recoiling target particle by $\theta_{T2}$, we find
\begin{equation}
    \cos\theta_{T2}=\frac{E_T E_2-m_T E_1+(m_2^2-m_1^2)/2}{\sqrt{(E_T^2-m_T^2)(E_2^2-m_2^2)}}\,. \label{eq:angseparation}
\end{equation}
This relation gives an idea of the angular separation between the recoiling target particle and the secondary $e^+e^-$ pair especially when the $\chi_2$ is significantly boosted, which is often the case for the electron scattering channel. 
The color-coded scattering plots in figure~\ref{fig:angcorr} show the unit-normalized relative event densities. 
The vertical axes are for the minimum of the angular separation between the target particle and the secondary electron and the angular separation between the target particle and the secondary positron, i.e., ${\rm Min}\left[\theta_{e_{\rm rec}^-e^-},\theta_{e_{\rm rec}^-e^+}\right]$ for the events with a recoiling electron and ${\rm Min}\left[\theta_{p_{\rm rec}e^-},\theta_{p_{\rm rec}e^+}\right]$ for the events with a recoiling proton. 
The top and bottom panels are respectively for the electron scattering case and the proton scattering case with benchmark reference points REF1 (left panels) and REF2 (right panels). 
For the electron channel, we see that the secondary electrons and positrons are close to the recoiling electron, whereas for the proton channel most of the recoiling protons roughly move in the direction orthogonal to the secondary particle momenta. 
Indeed, eq.~\eqref{eq:angseparation} predicts $\theta_{T2}=2.2^\circ$ (electron scattering, REF1), $3.8^\circ$ (electron scattering, REF2), $87^\circ$ (proton scattering, REF1), and $92^\circ$ (proton scattering, REF2) with a representative recoil electron energy of 200~MeV and 100~MeV, and a representative recoil proton kinetic energy being 21~MeV and 1.5~MeV, respectively.
These values are in good agreement with the corresponding values in figure~\ref{fig:angcorr}.
Finally, we show the angular separation between the secondary electron and positron on the horizontal axes in the plots of figure~\ref{fig:angcorr}.
For most of the events, these are very close to each other. 

Given that the angular distances between the final-state particles are small and can be smaller than the angular resolution of the DUNE detector, it is crucial to devise an efficient selection strategy in order to enhance the signal acceptance, which is an important part of this study.
As discussed earlier, the secondary decay vertex may be displaced from the primary scattering vertex, depending on the benchmark parameters of interest.
Two separated correlated vertices, i.e., resulting from the same event, 
allow for rejection of most background events.
We will define these selection criteria in section~\ref{sec:selection}. 
Another handle to suppress background is a ``$dE/dx$'' analysis.
This is inspired by the discrimination between electron-like tracks and photon-like tracks via $dE/dx$ performed by the ArgoNeuT Collaboration~\cite{Acciarri:2016sli}.
If charged particles are merged together, the $dE/dx$ 
measurement can be  distinctive enough to distinguish between signal and background events.
We will elaborate this idea in detail in section~\ref{sec:dedxanalysis}.

\section{DUNE detectors and backgrounds \label{sec:detector}}

We discuss the potential backgrounds  given the $i$BDM signature characteristics
discussed in the previous section. 
An $i$BDM signal consists of three visible particles in the final state potentially with a displaced vertex, which is generally 
difficult for SM processes (mostly by atmospheric neutrinos) to mimic.
We examine several scenarios, taking into consideration the anticipated detector performance such as spatial and angular resolutions, particle identification, etc. We first give an overview of the DUNE far detector, followed by a discussion of the backgrounds.

\subsection{DUNE far detectors}

DUNE is expected to start its initial operation in 2026/2027, carrying out various physics analyses including the measurement of neutrino oscillations, searches 
for proton decay, supernova neutrino observation, and more. 
Recently, DUNE has received increasing attention as an opportunity to probe new physics, such as detection of dark matter~\cite{Necib:2016aez, Alhazmi:2016qcs, Kim:2016zjx,BDDM,Kim:2020ipj,Berger:2019ttc}. 
While the experiment shares many common physics goals with similar experiments such as Hyper-Kamiokande, a different detector technology is adopted by DUNE and the other experiments so that a large degree of complementarity among the 
various experiments is expected.

The DUNE far detectors are designed to consist of four large-volume detectors with a fiducial volume of approximately 10\,kt each.
The main technology for at least three of these large detectors is that of a Liquid Argon Time Projection Chamber (LArTPC) for which two different complementary technologies are proposed: the single-phase (SP) and the dual-phase (DP) one.
The first module is scheduled to be ready and start collection of cosmic-ray and atmospheric neutrino data in 2026, two modules (i.e., a total of 20\,kt) are planned to be operational in 2027 with at that time also the FNAL neutrino beam turned on, and ultimately the full set of modules (i.e., the full 40 kt) are expected to be ready for operations three years later.

\begin{table}[t]
\centering
\scalebox{0.92}{
\begin{tabular}{c|c|c}
 \hline \hline
\multicolumn{2}{c|}{Target \& detector technology} & Liquid Argon \& LArTPC \\
\hline
\multicolumn{2}{c|}{Depth [m.w.e.]} & 4,300  \\
\hline 
\multirow{4}{*}{Dimension [m]} &\multirow{3}{*}{Active} & Cubic (${\rm width}\times{\rm length}\times{\rm height}$)  \\
 &  & SP: $14.0 \times 58.2 \times 12.0$ ($\times 2$)  \\
  &  & DP: $12.0 \times 62.0 \times 12.0$ ($\times 2$)  \\
\cline{2-3}
  & Fiducial$^\dagger$ & $11.2 \times 57.2 \times 11.2$ ($\times 4$) \\
\hline
\multirow{2}{*}{Mass [kt]} & Active & SP: $13.7\times 2$, DP: $12.1\times 2$  \\
\cline{2-3}
 & Fiducial &SP: $10.0\times 2$, DP: $10.0\times 2$  \\
\hline 
\multirow{2}{*}{$E_{\rm th}$ [MeV]} & electron & 30  \\
\cline{2-3}
 & proton & 30-50  \\
\hline
\multirow{5}{*}{$E_{\rm res}$ [\%]} & \multirow{3}{*}{electron} & 20 for $E<0.4$ GeV \\ 
&                         & 10 for $E<1.0$ GeV \\
&                         & $2+\frac{8}{\sqrt{E/{\rm GeV}}}$ for $E\geq 1.0$ GeV \\
\cline{2-3}
 & \multirow{2}{*}{proton} & 10 for $E<1.0$ GeV \\
 &                         & $5+\frac{5}{\sqrt{E/{\rm GeV}}}$ for $E\geq 1.0$ GeV \\
\hline
\multirow{2}{*}{$\theta_{\rm res}$ [$^\circ	$]} & electron & 1 \\ \cline{2-3}
 & proton & 5 \\
\hline
\multicolumn{2}{c|}{Vertex resolution $V_{\rm res}$ [cm]} & 1 \\
\hline \hline
\end{tabular}
}
\caption{\label{tab:detectors} A summary of the characteristics of a far detector similar to the one proposed by the DUNE Collaboration~\cite{Abi:2020wmh,Abi:2020evt,Abi:2020loh,Abi:2018rgm}.
The unit for depth, m.w.e., stands for meter-water-equivalent.
The ``$\dagger$'' symbol indicates  the quoted dimensions of the fiducial volumes used in section~\ref{sec:results} and
detailed in the text.
}
\end{table}

We summarize key specifications of the DUNE far detectors~\cite{Abi:2020wmh,Abi:2020evt,Abi:2020loh,Abi:2018rgm} in table~\ref{tab:detectors}.
In addition, pion tagging will be particularly important since it is deeply connected to rejection of potential backgrounds, which we will discuss in the next subsection. 
Several of the numbers in the table are taken from the latest technical design report (TDR)~\cite{Abi:2020wmh,Abi:2020evt,Abi:2020loh,Abi:2018rgm}, but a few comments are made in order.
First, the quoted dimensions of the fiducial volumes are the ones that we have defined for this data analysis discussed in section~\ref{sec:results}, following rough guidelines of the experiment.
We determine the fiducial volumes of both SP and DP detectors by removing at least $40-50$~cm
inward from the boundary of their active volumes, 
taking into account the modular readout plane structure.
Second, the energy threshold for electrons quoted in the table is the one used in the analyses for physics beyond the SM of the TDR~\cite{Abi:2020evt}, although 
other physics analyses, e.g., solar neutrino and supernova neutrino detection
suggest that a smaller value, as low as 5\,MeV may be possible~\cite{Abi:2020wmh,Abi:2020evt}. 
Third, the energy threshold for protons is (conservatively) estimated to be 50\,MeV~\cite{Abi:2020wmh,Abi:2020evt}, but the possibility of lowering it further in LArTPCs was discussed in ref.~\cite{Necib:2016aez}. 
As will be shown in section~\ref{sec:results}, the proton scattering channel particularly has a great potential in the search for dark-matter signal.
The ArgoNeuT Collaboration presented their study in ref.~\cite{Acciarri:2016sli} with a proton energy threshold in the LArTPC detector down to 21\,MeV.
However, the detailed design and granularity of the planned DUNE far detectors differ from those of ArgoNeuT.
Therefore, we take a somewhat less conservative value $E_{\rm th}=30$\,MeV for protons than that in the TDR as the baseline threshold value in our analyses as a compromise, in order to demonstrate the full power of the proton scattering channel.
Fourth, the precise numbers of the energy resolution in the region of interest here is still work in progress in the DUNE Collaboration beyond the references given.\footnote{The quoted numbers are based on private communications with members of the DUNE Collaboration.}
The numbers given in table~\ref{tab:detectors} are inspired by the expected DUNE detector performance but should not be taken as official numbers by the DUNE Collaboration.
In this analysis, we do not explicitly consider the event triggering, but DUNE foresees several low energy triggers for its physics programs such as supernovae detection, solar neutrinos, etc.

\subsection{Background consideration \label{sec:bkgd}}

As mentioned earlier, it is not easy for SM processes to mimic the $i$BDM-like signature in our study as depicted in figure~\ref{fig:signature}.
Since the DUNE far detectors will be placed deep underground, the background contamination from cosmic rays (mostly cosmic muons) is expected to be small. Nevertheless, the annual flux is not negligible, so a more detailed estimate has to be made.

The total muon flux at the DUNE detector location is $\sim 4\times10^{-5}~{\rm m}^{-2} {\rm sr}^{-1} {\rm s}^{-1}$~\cite{Agashe:2014kda}, resulting in $(1-2) \times 10^7$ muons annually at DUNE-40 kt.
The most plausible scenario for cosmic muon background is the following: the muon could sneak into the fiducial volume and emit a hard photon that converts into a $e^+e^-$ pair, and simultaneously leave either an electron-like or a proton-like track signature.
Although we need a more dedicated study on the probability of externally produced muons entering the DUNE far detector but not indentified as such, we can expect the probability to be less than 0.1\% from a study of the muon reconstruction efficiency at the MicroBooNE detector~\cite{MicroBooNENote}.\footnote{The MicroBooNE Collaboration reported that $0.09\%$ of cosmic muons are reconstructed such that their tracks appear only inside the fiducial volume~\cite{MicroBooNENote}.
While the value $0.09\%$ resulted from 2016 data of the MicroBooNE detector, the corresponding value including 2017 data is even smaller, although not public yet~\cite{private-MicroBooNE}.
We take this value as the upper limit on the probability of ``sneaking-in'' muon, and thus conservatively estimate the probability to be $0.1\%$.}
The rate of hard photon emission is suppressed by a factor of $\alpha/\pi\approx 1/500$, with $\alpha$ being the electroweak fine structure constant, and we estimate that the rate of electron-like muon tracks\footnote{The rate of proton-like muon tracks is much smaller.} is reduced by a conservative suppression factor of $10^{-2}$ based on the study in ref.~\cite{Acciarri:2016sli}.
Combining all factors together, we expect $\ll 1$ cosmic muon-induced background events per year at DUNE-40 kt.
The suppression factors are estimated very conservatively, anticipating that in reality the suppression power will be larger, but these need to be demonstrated with dedicated studies in the DUNE far detector.

The neutrinos coming from the sky may give rise to background events.
Atmospheric neutrinos can lead to a resonance scattering or a deep inelastic scattering (DIS) process, creating a handful of mesons (usually pions) whose visible decay products can leave signal-like signatures in a detector.
We expect that $\nu_e/\bar{\nu}_e$-induced charged-current events may mimic signal events.
For the electron channel, we can have for example: 
\begin{equation}
    \nu_e/\bar{\nu}_e + N \to e^\pm \pi^0 + N'\,,
\end{equation}
the $N$ and $N'$ are nuclei, and one of the two photons from the $\pi^0$ decay converts and appears electron-like.
Another class of the signal-looking example processes is
\bea
\nu_e/\bar{\nu}_e+ p\hbox{ or }n \rightarrow e^{\pm} \pi^{\pm}\pi^{\pm}+{\rm others}\,,
\eea
where ``others'' are sufficiently soft, i.e., undetected particles, and where charged pions themselves are misidentified as electrons.
There are other sources such as $\nu_\mu$-induced charged current events and $\nu$ neutral-current events.
The former are usually accompanied by an energetic muon which can be tagged very easily. 
The neutral-current contributions are subdominant typically measured to be $\sim 10-50\%$ of the corresponding charged-current contributions, depending on channel, energy, and target material~\cite{Formaggio:2013kya}.
So, we will focus on the $\nu_e/\bar{\nu}_e$ charged-current events only.  

\begin{figure}[t]
\centering
\includegraphics[width=7.3cm]{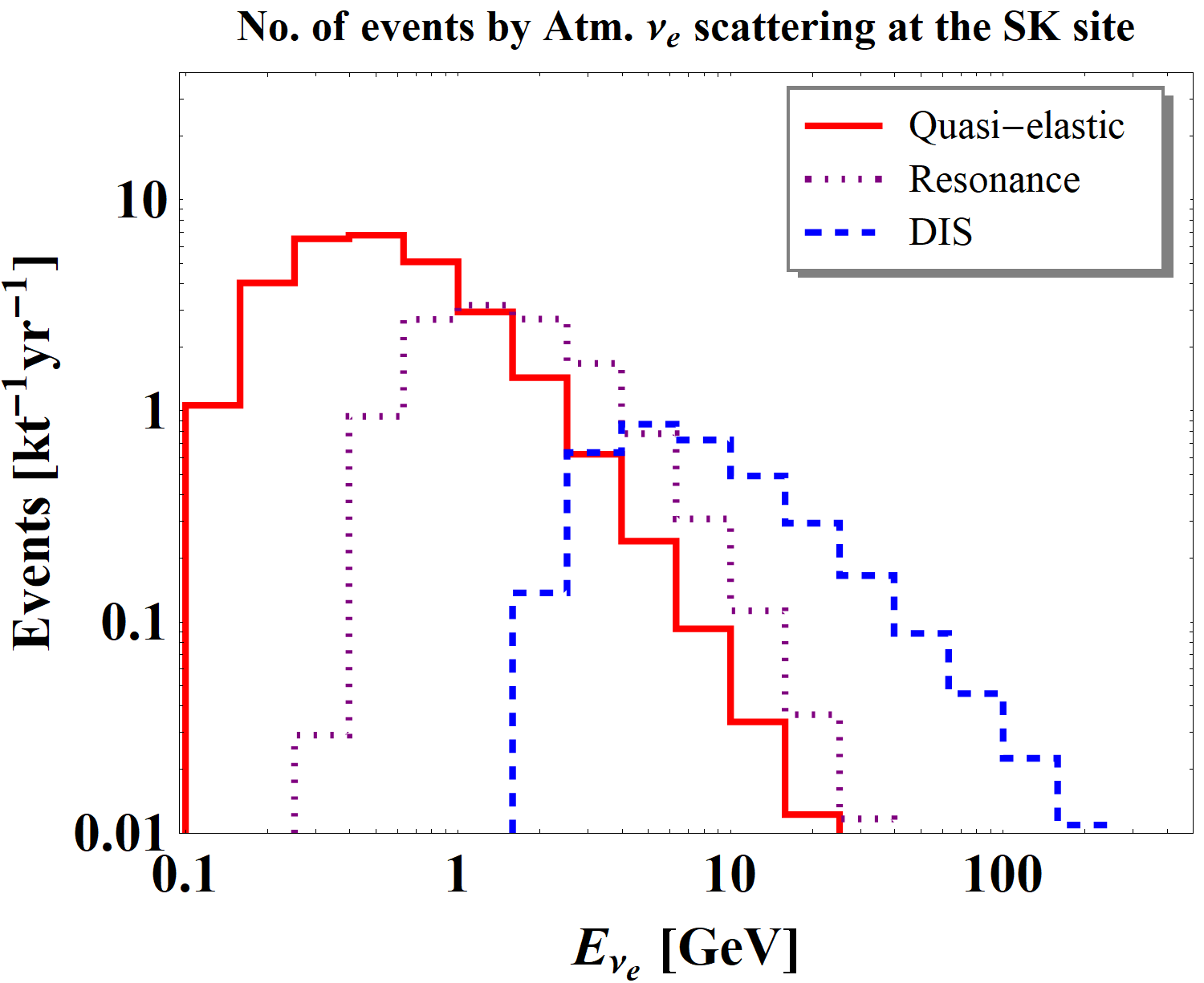} \hspace{0.2cm} 
\includegraphics[width=7.3cm]{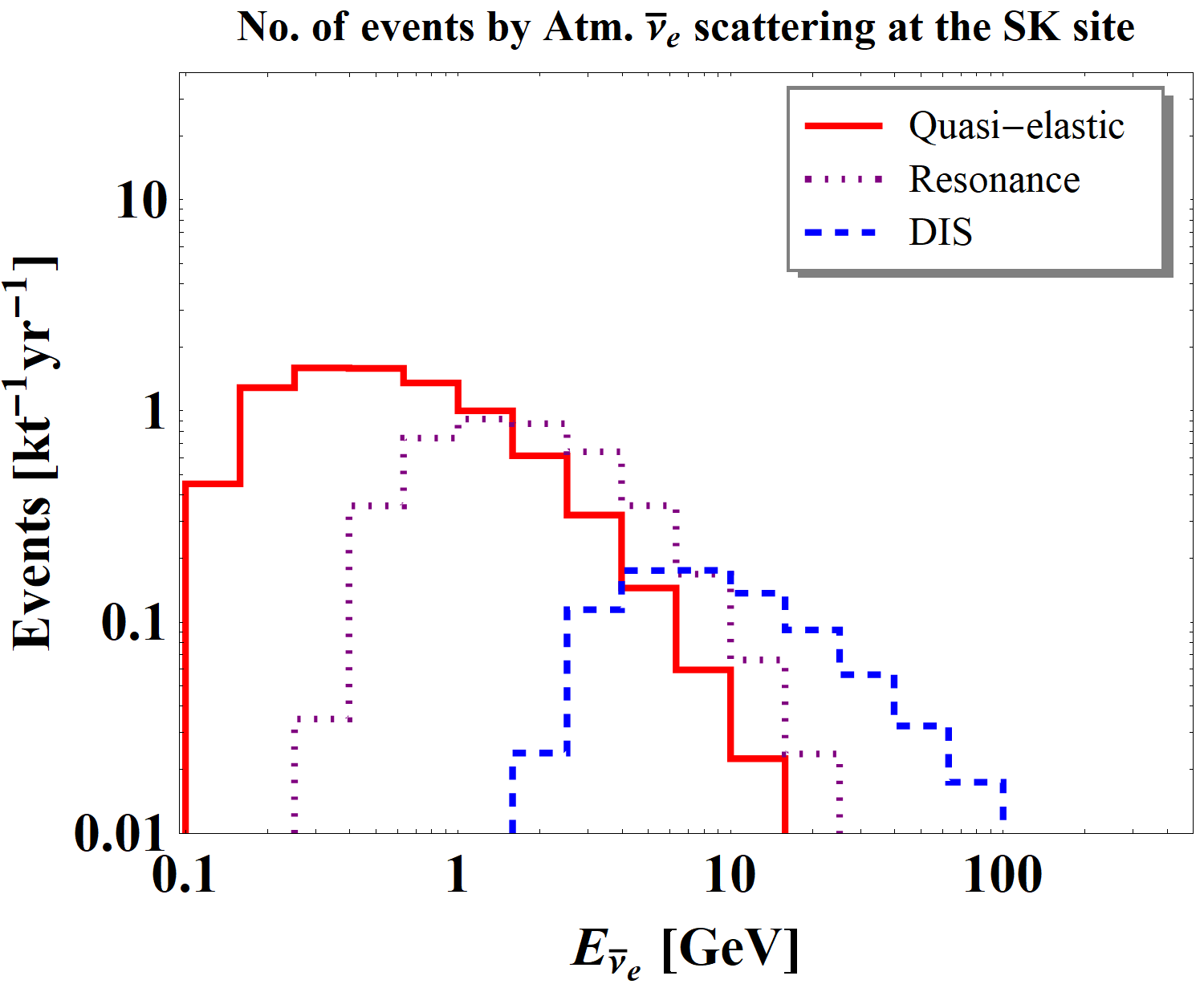}
\caption{\label{fig:nuflux} The annually expected number of atmospheric $\nu_e/\bar{\nu}_e$-induced events at a multi-kt-size detector. 
The left and right panels are for electron neutrino and anti-electron neutrino interactions, respectively.
The numbers are reported for each process: quasi-elastic in solid red, resonance in dotted purple, and DIS in dashed blue.}
\end{figure}

To estimate the rate of these events, we combine the atmospheric $\nu_e/\bar{\nu}_e$ differential flux for the Super-Kamiokande site as calculated in ref.~\cite{Honda:2015fha} and the neutrino scattering cross sections from ref.~\cite{Formaggio:2013kya}, and derive the number of events for the contributing channels according to the energy of the incident neutrino, as shown in figure~\ref{fig:nuflux}.
As the DUNE far detector site and the SK site are on similar latitudes and at similar depths, this is a good approximation.
The left and right panels are for electron neutrino and anti-electron neutrino scattering, respectively. 
We show the contributions from quasi-elastic scattering in solid red, resonance scattering in dotted purple, and DIS in dashed blue, for an exposure corresponding to one kt$\cdot$yr. 

We count all resonance scattering and DIS events in this energy range from figure~\ref{fig:nuflux}, and find that in total about 20 events are expected per kt$\cdot$yr. 
Therefore, in the full DUNE-40\,kt detectors, about 800 $\nu_e/\bar{\nu}_e$-induced events can potentially mimic a signal for a one-year exposure, whereby energy thresholds have been ignored for this simple estimate.
However, the events of this type usually contribute to the background when produced mesons and/or their decay products are not detected or incorrectly tagged.
Note that LArTPC detectors are expected to have good particle identification.
For example, the MicroBooNE Collaboration used a convolutional neural network to distinguish $\pi^\pm$ signatures from others in their LArTPC detector, and reported $70-75$\% $\pi^\pm$-tagging efficiency~\cite{Acciarri:2016ryt}.
This tagging efficiency is mainly to separate against $\mu^\pm$, and is much larger to separate against $e^\pm$.
We expect that a similar or better level of particle tagging efficiencies will be possible in the DUNE LArTPC detectors, and that such background events will be suppressed enough to be negligible. 

Another potential background is quasi-elastic scattering events of atmospheric electron-neutrinos (solid red histograms in figure~\ref{fig:nuflux}) involving a soft nucleon or a nucleus.
Since the nucleus or the (soft) nucleon recoil in such an event is invisible due to the energy threshold, only the $e^\pm$ will be visible. 
As will be elaborated in section~\ref{sec:dedxanalysis}, a $dE/dx$ cut may misidentify a certain fraction of single $e^\pm$ events as signal ones.
We count all quasi-elastic scattering events using the plots of figure~\ref{fig:nuflux} for a conservative estimate and find that about 37 events are expected per kt$\cdot$yr leading to $\sim1,500$ $\nu_e/\bar{\nu}_e$-induced quasi-elastic scattering events for the DUNE-40~kt detector with a one-year exposure, again ignoring energy thresholds.
Depending on the choice of $dE/dx$ cut, which will be discussed later, these events can be suppressed by $2-3$ orders of magnitude.
Moreover, some events involve a detectable proton recoil and can be recognized.
We therefore expect conservatively at most $\mathcal{O}(10)$ background events from this channel.

Finally, we comment briefly on the potential background events induced by beam-produced neutrinos and by random coincidences of events.
For the beam-produced neutrinos, the predicted event rates are $\sim2000$~yr$^{-1}$ in the neutrino mode and $\sim 800$~yr$^{-1}$ in the antineutrino mode for a 40~kt detector and  with $\delta_{\rm CP}=0$~\cite{Abi:2020evt}. 
In addition, the timing of such events is fully correlated with the neutrino beam bunch timing, and the visible particle tracks will, in general, be pointing back to the beam production source. 
Therefore, we expect that the beam-induced neutrino background can be safely eliminated.
For possible random coincidences, the rate was estimated to be negligible even in the surface-based ProtoDUNE detectors~\cite{Chatterjee:2018mej} and we therefore expect that the number of such background events can be neglected in the DUNE far detectors which will be placed deep underground.

\section{Event selection \label{sec:select}}

In this section, we discuss event selection scenarios used for our sensitivity studies.
Since boosted dark matter collides with a fixed target particle, the final-state particles are generally produced in the forward direction, i.e., following the incident $\chi_1$ direction.
In particular, in the electron scattering channel, a large boost factor is essential to produce the heavier dark-sector state.
As a result, all three electron final-state tracks may be highly collimated, and in some cases appear as a single electron track.
We first discuss a possible way of recognizing such a multi-track object.

\subsection{Identification of merged-track signal \label{sec:dedxanalysis}}

Electrons traveling in liquid argon lose their energy initially by ionization, before eventually developing an electromagnetic shower.
The radiation length in liquid argon is 14 cm, so the first few centimeters of the track, before the electron starts showering, are generally relatively clean and allow for an accurate measurement of the ionization energy depositions per unit length, i.e., the $dE/dx$.
This quantity is a characteristic of particles moving in material, and depends on the particle mass, its velocity and the material parameters, and can be utilized as a metric to identify particles.
For example, an energetic photon converts into an electron-positron pair in the liquid argon, and the two tracks are likely to be close by and overlaid, and may be reconstructed as a single electron track.
The DUNE far detectors do not have a magnetic field and so cannot separate electrons and positrons using track curvature.
Being in fact the sum of two tracks, on average the $dE/dx$ value of this $\gamma$-induced ``track'' will be twice as large as that of a single electron track.  

\begin{figure}[t]
    \centering
    \includegraphics[width=7.3cm]{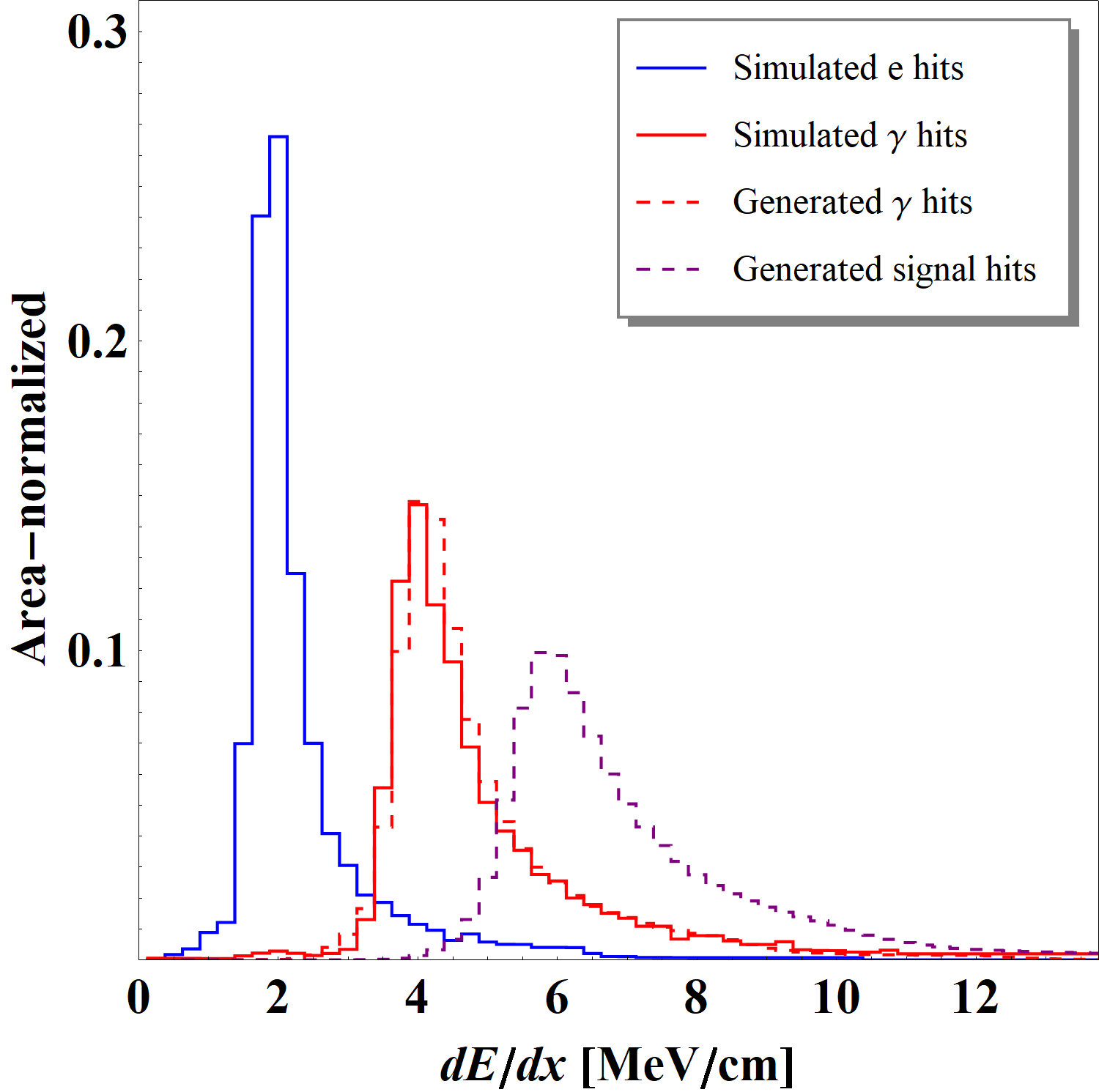} \hspace{0.2cm}
    \includegraphics[width=7.3cm]{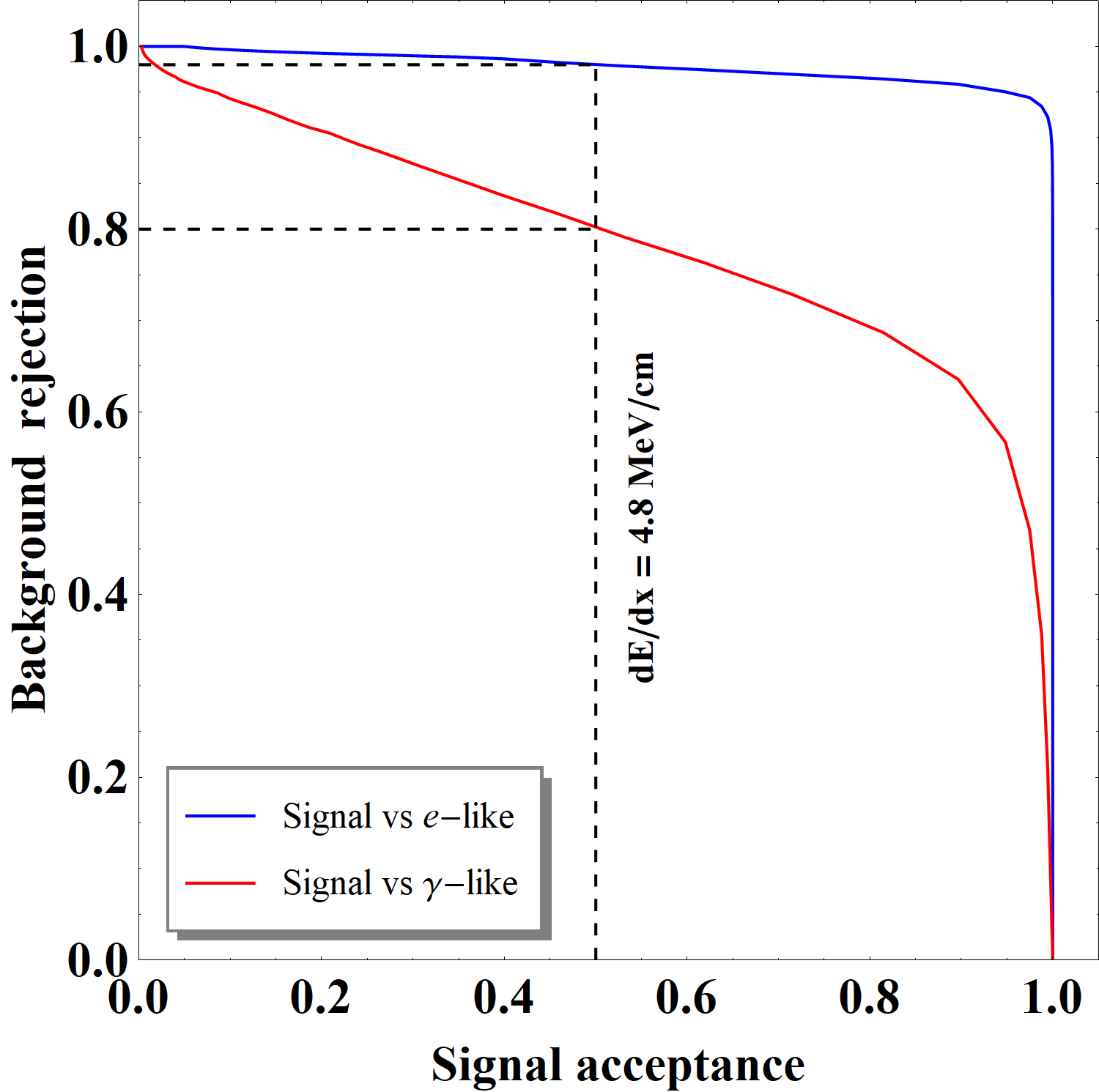}
    \caption{\label{fig:dedx} Left: $dE/dx$ distributions of electron events (solid blue) and photon events (solid red) simulated by the DUNE Collaboration with the default detector design~\cite{Abi:2020evt}.
    The dashed red histogram shows $\gamma$ hits generated using the simulated $e$ hits (solid blue) of DUNE, while the solid red one shows the simulated $\gamma$ hits.
    The dashed purple histogram is a prediction for signal events (i.e., three-electron merged tracks) generated using the simulated $e$ hits. 
    Right: ROC curves of signal acceptance versus background rejection.
    The blue curve compares the generated signal hits with the simulated $e$ hits, whereas the red curve compares the generated signal hits with the simulated $\gamma$ hits.
    The dashed lines correspond to 50\% of signal acceptance, $\sim 80$\% of $\gamma$-like event rejection, and $\sim 98$\% of single $e$-like event rejection with a $dE/dx$ cut imposed at 4.8~MeV/cm.}
\end{figure}

This effect forms the basis of the strategy that the ArgoNeuT Collaboration has taken to distinguish electron-induced tracks from photon-induced tracks in their detector~\cite{Acciarri:2016sli}.
They observed that the electron hits follow a Gaussian convolved with a Landau spectrum peaking at $dE/dx\approx 2$~MeV/cm, while the spectrum for the $\gamma$ hits shows a peak at $dE/dx\approx 4$~MeV/cm.
The DUNE Collaboration has performed such a study with simulated data and reached a similar conclusion~\cite{Abi:2020evt}, using samples of electron and photon electromagnetic cascades with isotropic directions and uniformly distributed momenta in the range 0.2 GeV to 5.0 GeV.
Therefore, we can view the probability density $P_{\rm hit}^\gamma$ associated with the $\gamma$ hits as a combination of two probability densities $P_{\rm hit}^e$ associated with the electron hits:
\begin{equation}
    P_{\rm hit}^\gamma(x)=\int_0^x dy P_{\rm hit}^e(x-y)P_{\rm hit}^e(y)\,, \label{eq:2hits}
\end{equation}
where the argument of $P_{\rm hit}$ is the  $dE/dx$ value.
This implies that one can generate $P_{\rm hit}^\gamma$ based on the knowledge of $P_{\rm hit}^e$.
In the left panel of figure~\ref{fig:dedx}.
We compare our own ``generated'' $\gamma$ hits with the $\gamma$ hits simulated by DUNE with the default detector design in ref.~\cite{Abi:2020evt}.
The solid blue histogram serves as the input $P_{\rm hit}^e$ from which we generate the dashed red histogram. 
Comparing this with the solid red histogram, we find that the generated hits reproduce the simulated hits fairly well. 

Inspired by this result, we extend eq.~\eqref{eq:2hits} to our signal in the 
most extreme case, i.e. the one where all three electron/positron 
tracks merge:
\begin{equation}
    P_{\rm hit}^{\rm sig}(x)=\int_0^x \int_0^{x-y} dy dz P_{\rm hit}^e(x-y-z)P_{\rm hit}^e(y) P_{\rm hit}^e(z) \,. \label{eq:3hits}
\end{equation}
This formula allows  to generate the signal hits and the prediction is shown by the purple dashed histogram in the left panel of figure~\ref{fig:dedx}. 
The generated signal hits predict the peak position ($\sim6$~MeV/cm) at around three times the peak of simulated electron hits or equivalently 1.5 times the peak of simulated $\gamma$ hits, as expected. 

We further study signal acceptance versus background rejection by comparing the generated signal hits with the simulated 
electron and photon hits, and represent
the comparisons as a Receive Operating Characteristic (ROC) curve. The right panel of figure~\ref{fig:dedx} shows two ROC curves, signal hits against electron hits (blue) and signal hits against photon hits (red). 
The dashed lines correspond to a $dE/dx$ cut at 4.8~MeV/cm which is roughly the crossover point between the signal hits and the photon hits, and this choice allows for 
50\% acceptance of signal-like events, $\sim80$\% rejection of $\gamma$-like events, and $\sim98$\% rejection of 
single electron events.  
We take 50\% as our baseline tagging efficiency for the three-electron/positron merged tracks in the selection criteria detailed in the next subsection.
Future studies with the well-tuned simulation and reconstruction tools in DUNE may validate this method further.

\subsection{Event simulation and selection criteria \label{sec:selection}}

We  discuss the event selection scheme for the sensitivity study  reported in section~\ref{sec:results}. 
An event is generated as follows. First, the primary scattering point of the dark-matter particle within the detector is generated randomly inside the fiducial volume of a single module of the DUNE far detector.
Second, for a given set of $E_1$, $m_1$, $m_2$, $m_X$, and $m_T$ (either $m_e$ or $m_p$) parameters, the four-momenta of the recoiling target particle and the produced $\chi_2$ are generated according to the associated recoil energy spectrum based on the appropriate scattering matrix element. 
Three-momentum directions are defined accordingly under the assumption that the yearly average of the incoming $\chi_1$ flux is isotropic. 
Third, the laboratory-frame lifetime of the long-lived particle (either $\chi_2$ or on-shell $X$) is calculated and a decay length is generated by a conventional  
exponential decay distribution.
Fourth, the secondary decay vertex position is calculated  using the decay length and the momentum of the long-lived particle.
Finally, the decay is generated, leading to  the  $e^+e^-$ decay products.

Once the event generation is completed, the following selection criteria are consecutively tested to determine if the event is accepted:
\begin{enumerate}
    \item {\bf Energy}: Energy of final-state protons and electrons is smeared according to the energy resolution formulas tabulated in table~\ref{tab:detectors}.
    If the resulting smeared energy does not meet the threshold requirement ($E_{{\rm th}}=30$~MeV for electrons and $E_{{\rm th}}^{\rm kin}=30$~MeV for protons), the event is rejected.
    We also require the energy of recoiling protons not to exceed 2~GeV beyond which deep inelastic scattering processes of $\chi_1$ become significant~\cite{Fechner:2009aa}.\footnote{For most of parameter choices and parameter space that we study in this paper, the energy of recoiling protons is much less than 2~GeV~\cite{Kim:2020ipj}, so the precise value of choice of 2~GeV has only a negligible effect on our analyses.}
    
    \item {\bf Track containment}: The track length of final-state particles are estimated, based on the expected electron and proton stopping power in liquid argon~\cite{ICRU:1984dou,ICRUBethesda:1994esa}.
    If the endpoint of a track lies outside the defined fiducial volume, the event is rejected. 
    
    \item {\bf Displaced vertex}: If the decay vertex falls outside the fiducial volume, the event is rejected.
    However, if it is displaced and the decay length is larger than the position resolution of 1~cm and the track length of recoiling particle, the event is accepted.
    Alternatively, if the decay length is between 1~cm and the track length of recoiling particle but the angular separation between the decay point and the recoiling particle is five times larger (smaller) than the angular resolution of recoiling particle (i.e., $5\time 1^\circ$ for electron recoil and $25\time 1^\circ$ for proton recoil), the event is accepted (rejected).
    
    \item {\bf Angular separation}: If the decay length is less than the position resolution 1~cm and the angular distances between pairs of final-state particles are greater than the corresponding angular resolutions ($\theta_{\rm res}=1^\circ$ for electrons and $\theta_{\rm res}=5^\circ$ for protons), the event is accepted.
    Otherwise, the event is identified as merged for the electron channel but it is rejected for the proton channel. 
    
    \item {\bf \textit{dE/dx}}: For the electron channel, if the event is identified as merged, it is accepted with an efficiency of 50\%.
\end{enumerate}
Note that these selection criteria are driven by the anticipated instrumental capabilities of the DUNE LArTPC detectors, and have not been optimized for the detailed signal event topology under consideration and for different search regions; some of them could be adapted to increase the sensitivity in some regions of parameter space.
With the study in this paper, we simply aim to demonstrate the huge potential of the DUNE far detectors for the search of cosmogenic new physics signals involving events with a multiple particle signature.

\section{Results \label{sec:results}}

In this section, we study expected sensitivities to the dark-matter signal depicted in figure~\ref{fig:signature}, using event simulation and the event selections described in the previous section. 
Since the benchmark model contains a dark photon $X$, it is natural to investigate the experimental sensitivity in the standard dark-photon parameter space, $\epsilon$ against $m_X$, for the DUNE far detectors. 
Models of the inelastic boosted dark-matter scenario contain more parameters, namely $m_0$, $m_1$, $m_2$, and $g_{12}$ in addition to these two.  
For definiteness, we take $g_{12}=1$ throughout this section whenever necessary and examine several different reference mass points, including REF1 and REF2 introduced in section~\ref{sec:kin}. 
These parameter choices are the same as those in ref.~\cite{Chatterjee:2018mej} which discussed similar sensitivities for the electron scattering channel using  the ProtoDUNE detectors, and have been used as reference parameter choices for the study reported in the DUNE TDR~\cite{Abi:2020evt}.

For a given time of exposure $t_{\rm exp}$ and number of target particles inside the detector fiducial volume $N_T$, the expected number of observed signal events $N_{\rm sig}$ is given by
\begin{equation}
    N_{\rm sig} = \sigma\ \mathcal{F}_1\  A_{\rm exp}\ t_{\rm exp}\ N_T\,, \label{eq:Nsignal}
\end{equation}
where $\sigma$ is the $\chi_1$ scattering cross section and $A_{\rm exp}$ stands for the experimental signal efficiency and acceptance following the event selection criteria in section~\ref{sec:selection}.
We take eq.~\eqref{eq:fluxform} for the flux factor $\mathcal{F}_1$ which was determined with $\chi_0$ following the NFW dark-matter halo profile; different $\chi_0$ halo profiles may lead to different values.
We further assume that yearly-averaged $\mathcal{F}_1$ is approximately isotropic.
The scattering cross section $\sigma$ for the $\chi_1T \to \chi_2T$ process can be obtained by integrating the differential cross section in eq.~\eqref{eq:diffscat} over the range defined by eq.~\eqref{eq:eTrange}.
The decay branching fraction of $\chi_2$ to $\chi_1e^-e^+$ is assumed unity for definiteness.

Next, sensitivity calculations are performed in both a model-dependent and a model-independent way.
For the former case, we investigate the sensitivity of DUNE to several representative model points in our benchmark model, 
both in the standard parameter space of dark-photon mass versus kinetic mixing parameter and in the plane of halo dark-matter mass $m_0$ versus the velocity-averaged annihilation cross section.
For the latter case, we discuss ways to present the results.

\subsection{Model-dependent sensitivity reaches \label{sec:modeldep}}

We consider the 90\% C.L. exclusion limits $N^{90}$ calculated with a modified frequentist construction~\cite{Read:2000ru,ATLAS:2011tau}.
An experiment is said to be sensitive to a given signal, if $N_{\rm sig}\geq N^{90}$.
The background estimation determines $N^{90}$.
Factoring out $\epsilon^2$ from the cross section, i.e., $\sigma=\epsilon^2 \tilde{\sigma}$ and substituting eq.~\eqref{eq:Nsignal} into this inequality, we have 
\begin{equation}
    \epsilon^2 \geq \frac{N^{90}}{\tilde{\sigma}(m_X)\  \mathcal{F}_1\ A_{\rm exp} \ t_{\rm exp} \ N_T}\,, \label{eq:senseeps}
\end{equation}
where the dependence of $\tilde{\sigma}$ on the mass of dark photon $X$ is explicitly shown. 
Therefore, if no additional events are observed beyond known backgrounds, any $\epsilon^2$ values greater than the value of the right-hand side of \eqref{eq:senseeps} are excluded for a given $m_X$.

\begin{figure}[t]
    \centering
    \includegraphics[width=7.3cm]{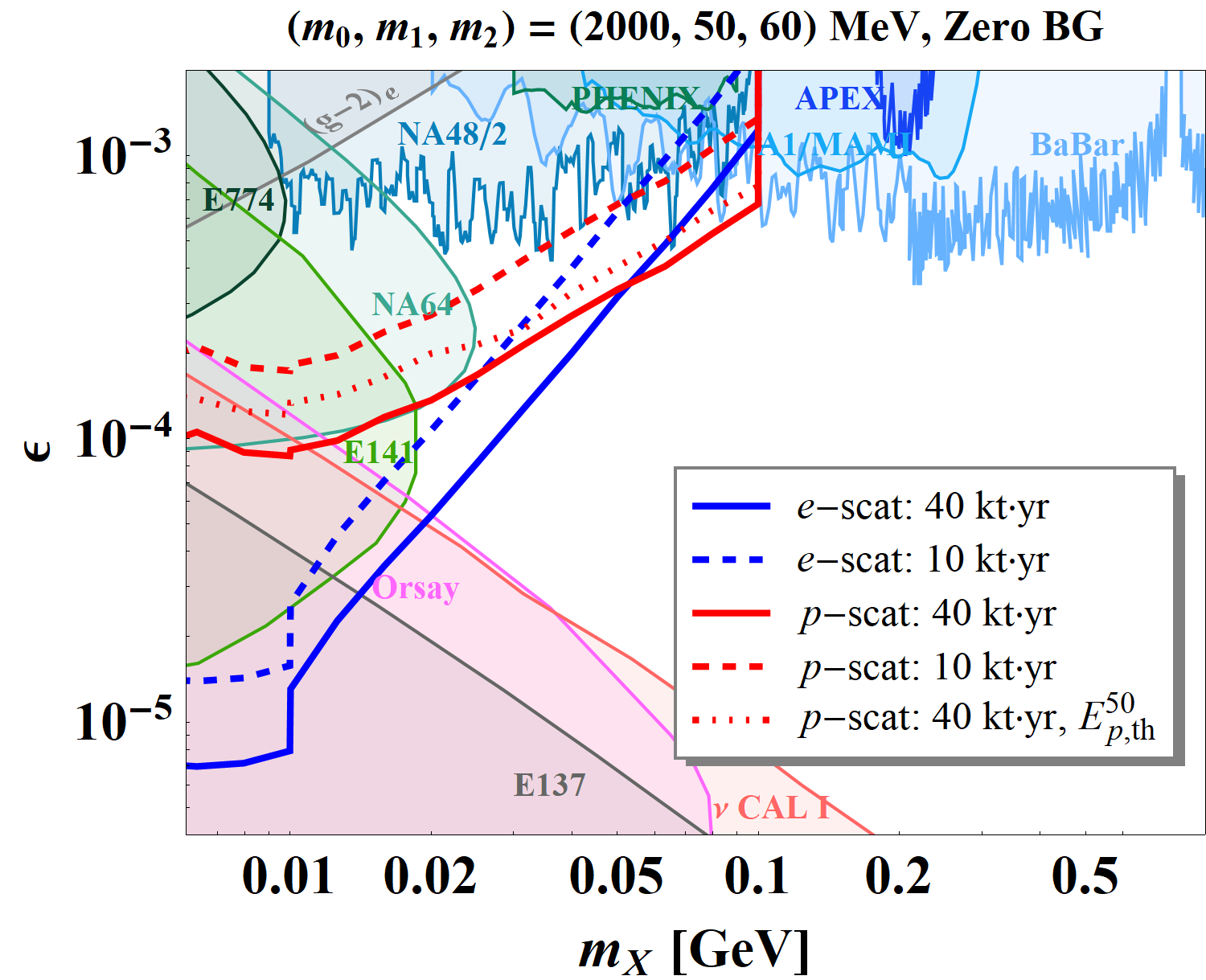} \hspace{0.2cm}
    \includegraphics[width=7.3cm]{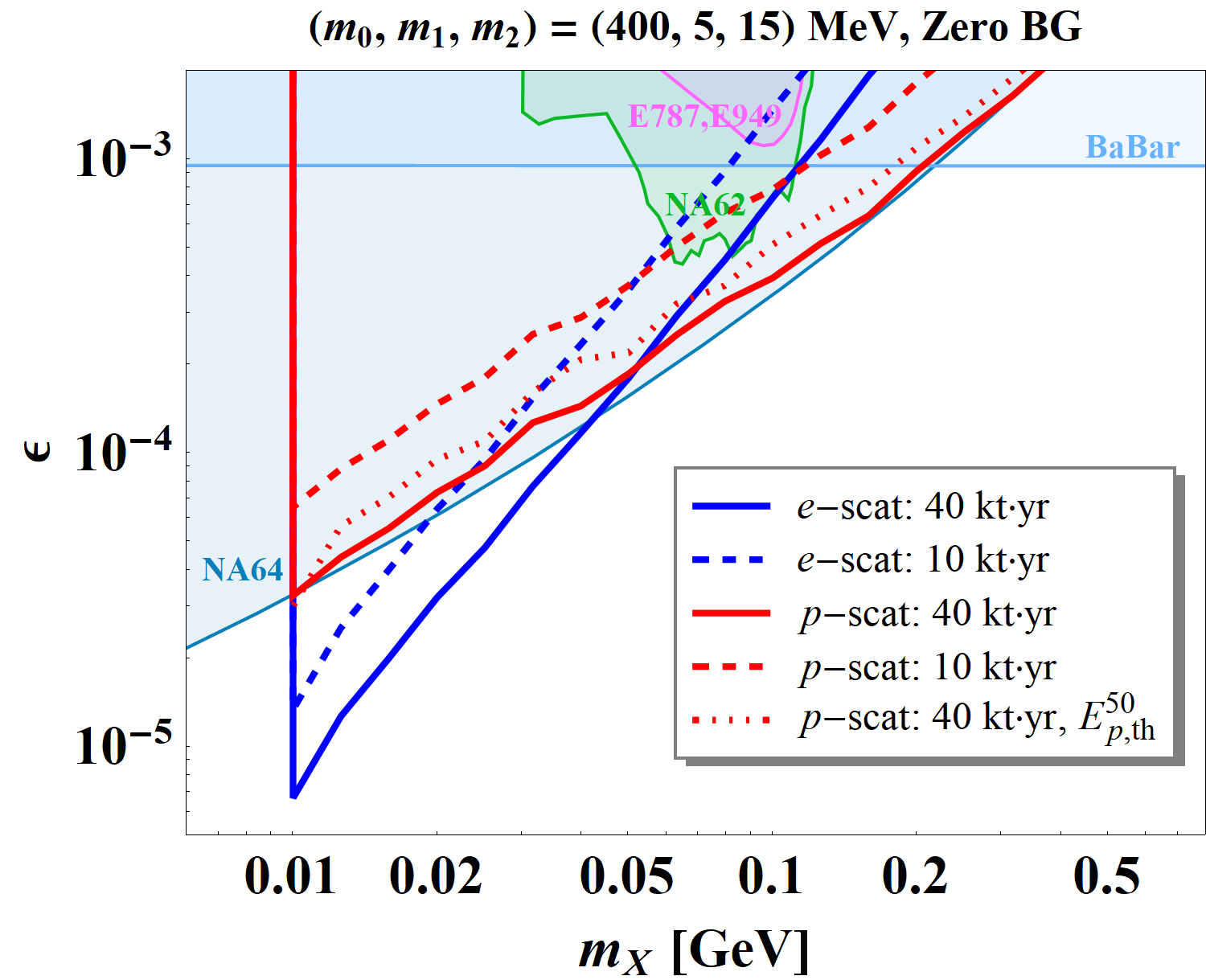} \\ \vspace{0.2cm}
    \includegraphics[width=7.3cm]{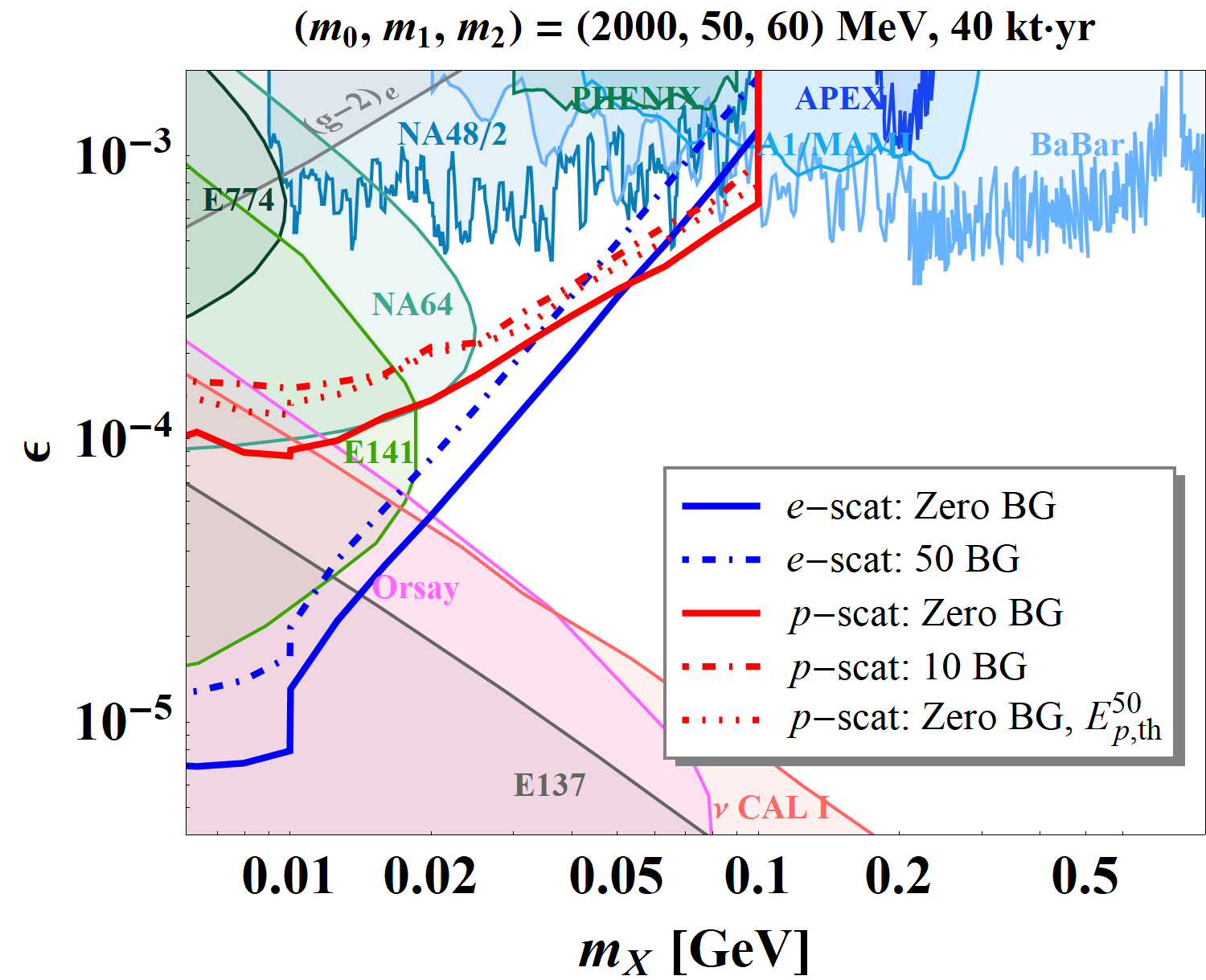} \hspace{0.2cm}
    \includegraphics[width=7.3cm]{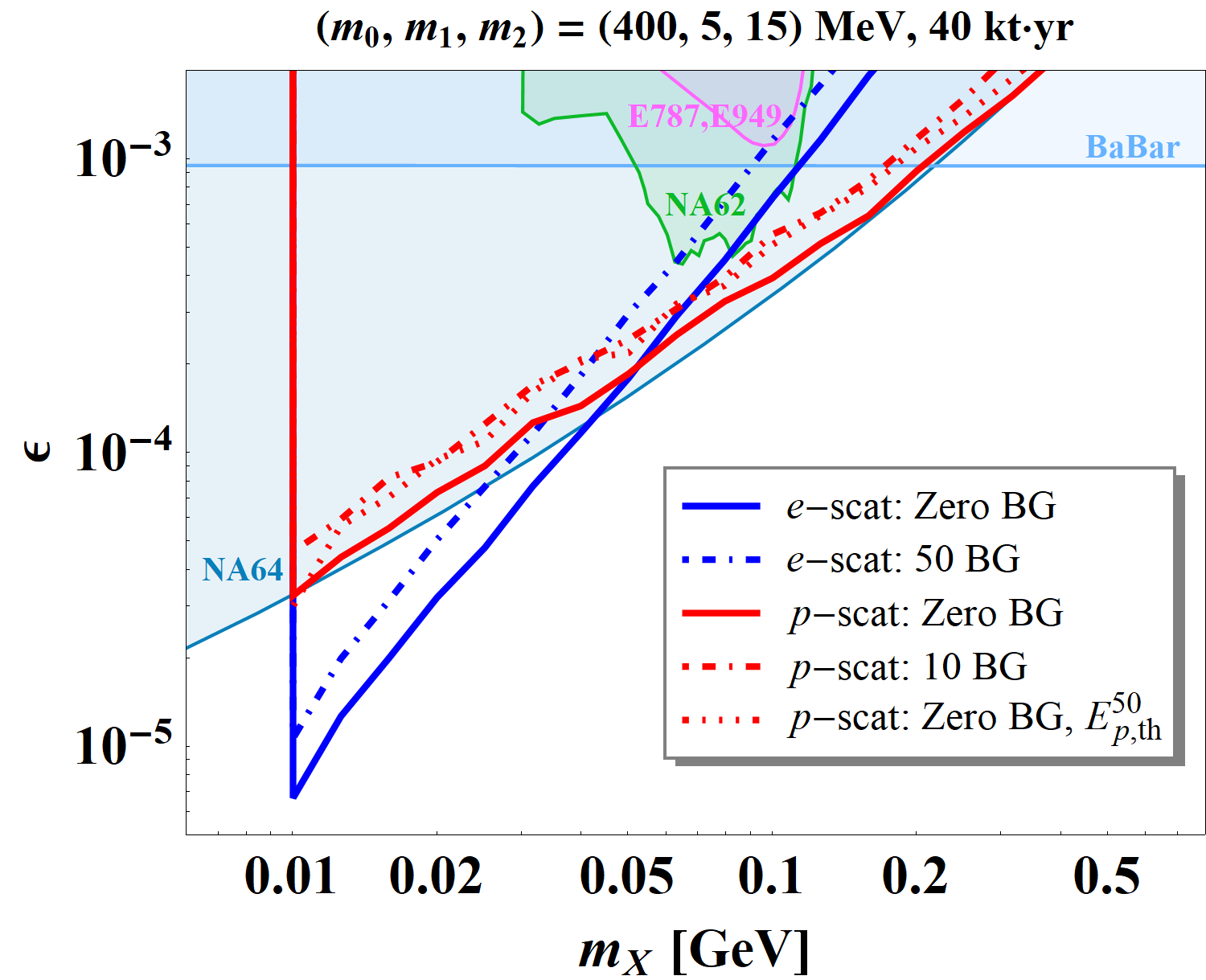} 
    \caption{Left: Experimental sensitivities of the electron scattering channel (blue lines) and the proton scattering channel (red lines) for REF1 for which $m_X$ is varied within the range of $m_X < 2m_1$.
    Relevant existing limits are taken from refs.~\cite{Riordan:1987aw,Bjorken:1988as,Davier:1989wz,Bross:1989mp,Blumlein:1990ay,Blumlein:1991xh,Abrahamyan:2011gv,Merkel:2014avp,Lees:2014xha,Adare:2014mgk,Batley:2015lha,Banerjee:2019hmi}.
    In the top panel, assuming negligible backgrounds, we compare the results with a statistics of 40~kt$\cdot$yr (solid lines) and  with a statistics of 10~kt$\cdot$yr (dashed lines).
    In the bottom panel, assuming a statistics of 40~kt$\cdot$yr, we compare the results with negligible backgrounds (solid lines) with the results with 50 and 10 background events (dot-dashed lines) for the electron and proton channel, respectively.
    For comparison, we show the result with $E_{{\rm th}}^{\rm kin}=50$~MeV for the proton channel (dotted line) in both panels.    Right: Experimental sensitivities to REF2 for which $m_X$ is varied within the range of $m_X>2m_1$.
    Relevant existing limits are taken from refs.~\cite{Davoudiasl:2014kua,Essig:2013vha,Lees:2017lec,CortinaGil:2019nuo,NA64:2019imj}.
    }
    \label{fig:epsvsmX}
\end{figure}

We now consider a few representative experimental scenarios.
As described in section~\ref{sec:detector}, one out of the four far detector modules will be ready at the start of the  data collection, so we will calculate sensitivities with DUNE 10~kt times one duty year (denoted by DUNE-10~kt$\cdot$yr) as well as with full detector DUNE 40~kt times one duty year (denoted by DUNE-40~kt$\cdot$yr).
Regarding the background, we consider not only an optimistic scenario of a negligible background level (denoted by Zero BG), for which $N^{90}({\rm Zero~BG})=2.3$, but also a more conservative scenario allowing for a sizable amount of background events.
According to the discussion in section~\ref{sec:bkgd}, several tens of atmospheric neutrino-induced background events could be selected in the electron scattering channel.
We therefore assume 50 events per 40 kt$\cdot$yr for a conservative scenario for the electron channel (denoted by 50 BG) for which $N^{90}({\rm 50~BG})=13.0$.
By contrast, since the proton scattering channel requires a detectable recoiling proton, only a tiny fraction of quasi-elastic scattering neutrino events and resonance events are expected to be selected as signal events and we take 10 events per 40 kt$\cdot$yr for the conservative scenario for the proton channel (denoted by 10 BG) for which $N^{90}({\rm 10~BG})=6.6$. 
Our sensitivity results on the $m_X-\epsilon$ plane are shown in figure~\ref{fig:epsvsmX}.
The left panels show the experimental sensitivities of the electron scattering channel (blue lines) and the proton scattering channels (red lines) for REF1 for which $m_X$ is varied within the range of $m_X<2m_1$, as well as existing experimental limits from refs.~\cite{Riordan:1987aw,Bjorken:1988as,Davier:1989wz,Bross:1989mp,Blumlein:1990ay,Blumlein:1991xh,Abrahamyan:2011gv,Merkel:2014avp,Lees:2014xha,Adare:2014mgk,Batley:2015lha,Banerjee:2019hmi}.
In contrast, the right panels show the corresponding experimental sensitivities for the benchmark point  REF2, for which $m_X$ is varied in the range of $m_X\geq 2m_1$, as well as various experimental limits from refs.~\cite{Davoudiasl:2014kua,Essig:2013vha,Lees:2017lec,CortinaGil:2019nuo,NA64:2019imj}.
The top panels compare the results with 40~kt$\cdot$yr (solid lines) and the results with 10~kt$\cdot$yr (dashed lines), assuming negligible backgrounds.
The bottom panels compare the results with Zero BG (solid lines) and the results with 50 BG/10 BG (dot-dashed lines), assuming  40~kt$\cdot$yr.
We also present the result with $E_{{\rm th}}^{\rm kin}=50$~MeV for the proton channel (dotted line) to show the dependence of the sensitivity reaches on $E_{{\rm th}}^{\rm kin}$.

We make several observations on these results.
First, we see that the electron scattering channel generally shows a better signal sensitivity than the corresponding proton scattering one for small mass values of $X$, but this trend is reversed as $m_X$ increases.
The $\chi_1$ scattering cross section on protons is larger than the one on electrons, see eq.~\eqref{eq:diffscat}.
However, if $m_1 \ll m_p$, the energy transfer to the target proton is not efficient so that the recoiling proton for a large fraction of events does not pass the energy threshold to be observed.
For the electron channel $m_1 \gg m_e$, and thus the recoiling electrons pass easier the observation threshold.
Eventually, this gets alleviated with increasing $m_X$, i.e., more recoiling protons lead to an energy deposit above the threshold and we have a crossover between the sensitivity curves of the electron and proton scattering channels (see also ref.~\cite{Kim:2020ipj} for a more systematic discussion).

Second, we find that taking into account the background assumptions
does not substantially degrade the signal sensitivities.
Indeed, the comparison between the top panels and the bottom panels of figure~\ref{fig:epsvsmX} suggests that experimental
exposure time be more important.
Finally, for REF2 the expected sensitivity reach of DUNE is slightly beyond the existing bound given by present NA64~\cite{NA64:2019imj} (the electron scattering case) or comparable to the bound (the proton scattering case).
NA64 will collect more data in the next years and will improve their sensitivity. But note that the search by NA64 assumes an invisible decay of the (on-shell) dark photon into a dark-matter pair ($X\to  \chi_1\bar{\chi}_1$), namely, a search 
based on ``disappearance'' signature.
Reversely, in our case the dark photon decays visibly through an off-shell intermediary state appearing in the $\chi_2$ decay process ($\chi_2\to \chi_1 X^* \to \chi_1 e^+e^-$), i.e., it is a search based on ``appearance'' signature.
Moreover, different choices of model points allow us to probe unexplored regions,\footnote{Needless to say, it is obvious that more data collection (say, 5-year duty run) improves experimental reaches.} which we will discuss shortly. 

Simulation studies were also performed with more conservative values of the angular and position resolutions.
These could potentially have an important impact on the sensitivity reach since the resolutions are closely related to the criteria for isolating individual particles and identifying a displaced vertex, and thus affect the signal acceptance.
We separately inflate the associated baseline values (i.e., $\theta_{\rm res}$ and $V_{\rm res}$) by a factor of 3, and find that the sensitivity curves reported in figure~\ref{fig:epsvsmX} are not significantly degraded by the variation of these deteriorated resolutions. 
As stated earlier, we also performed our simulation studies with a more conservative proton energy threshold of $E_{{\rm th}}^{\rm kin}=50$~MeV assuming negligible backgrounds per 40 kt$\cdot$yr. 
The results show similar sensitivity reaches to the cases with $E_{{\rm th}}^{\rm kin}=30$~MeV assuming 10 background events per 40 kt$\cdot$yr or negligible backgrounds per 20 kt$\cdot$yr.
These studies show that we can reduce the required time exposure by half by improving the energy threshold for detecting protons from 50 MeV to 30 MeV.

\begin{figure}[t]
    \centering
    \includegraphics[width=7.3cm]{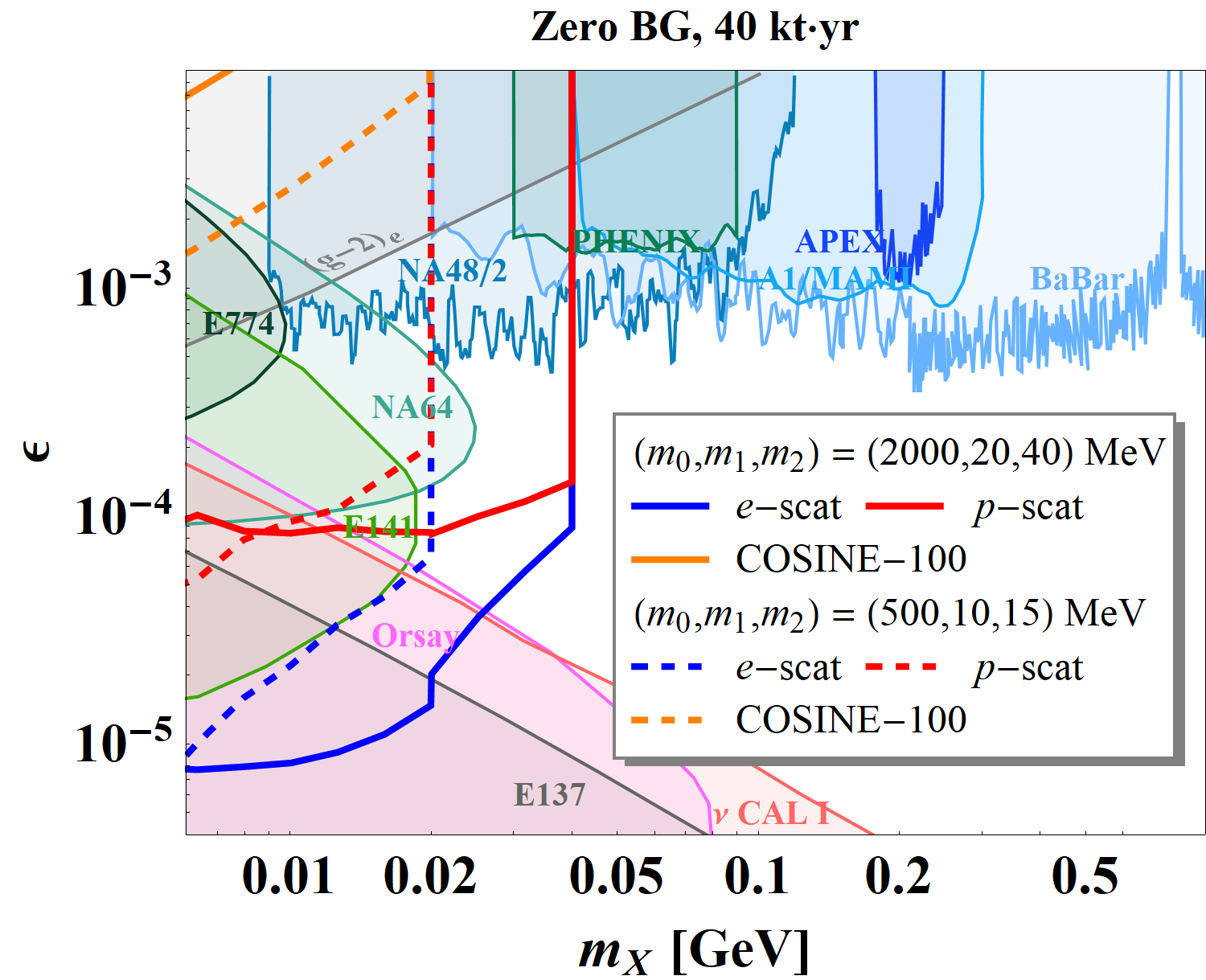}
    \caption{Comparison between the sensitivities of DUNE and COSINE-100 for the two benchmark points investigated by the COSINE-100 Collaboration~\cite{Ha:2018obm}. Parameter values and line specifications are detained in the legend. For the DUNE sensitivities, negligible background and 40~kt$\cdot$yr data collection are assumed.}
    \label{fig:cosine}
\end{figure}

It is informative to compare the experimental sensitivity reaches with the ones reported by the COSINE-100 Collaboration, as they performed the first $i$BDM signal search~\cite{Ha:2018obm}.
Three benchmark points were investigated.
However, $E_1$ of a point, for which COSINE-100 probed a new region beyond the existing bounds, is too small for (some of) the associated final state visible particles to overcome the energy thresholds of DUNE.
So, we compare the remaining two points in figure~\ref{fig:cosine} where the reported COSINE-100 sensitivities are shown by the orange solid and the orange dashed lines.
The corresponding sensitivity reaches are shown by the blue/red solid and the blue/red dashed curves in the electron/proton scattering channel, respectively, with the assumptions of negligible background and an exposure of 40 kt$\cdot$yr. 
For the first benchmark point with $m_0=2$~GeV (solid lines), the expected flux of $\chi_1$ is too small for COSINE-100 to cover a wide range of parameter space, especially toward smaller $\epsilon$, compared to its detector volume, whereas DUNE enjoys its large detector volume and is expected to achieve better sensitivity reaches as also advocated by the results in the left panel of figure~\ref{fig:epsvsmX}.
For the other benchmark point with $m_0=0.5$~GeV (dashed lines), the flux of boosted $\chi_1$ increases, resulting in an improved sensitivity reach of COSINE-100 within the excluded regions, while DUNE still would be able to probe some of the unexplored regions. 

\begin{figure}[t]
    \centering
    \includegraphics[width=7.3cm]{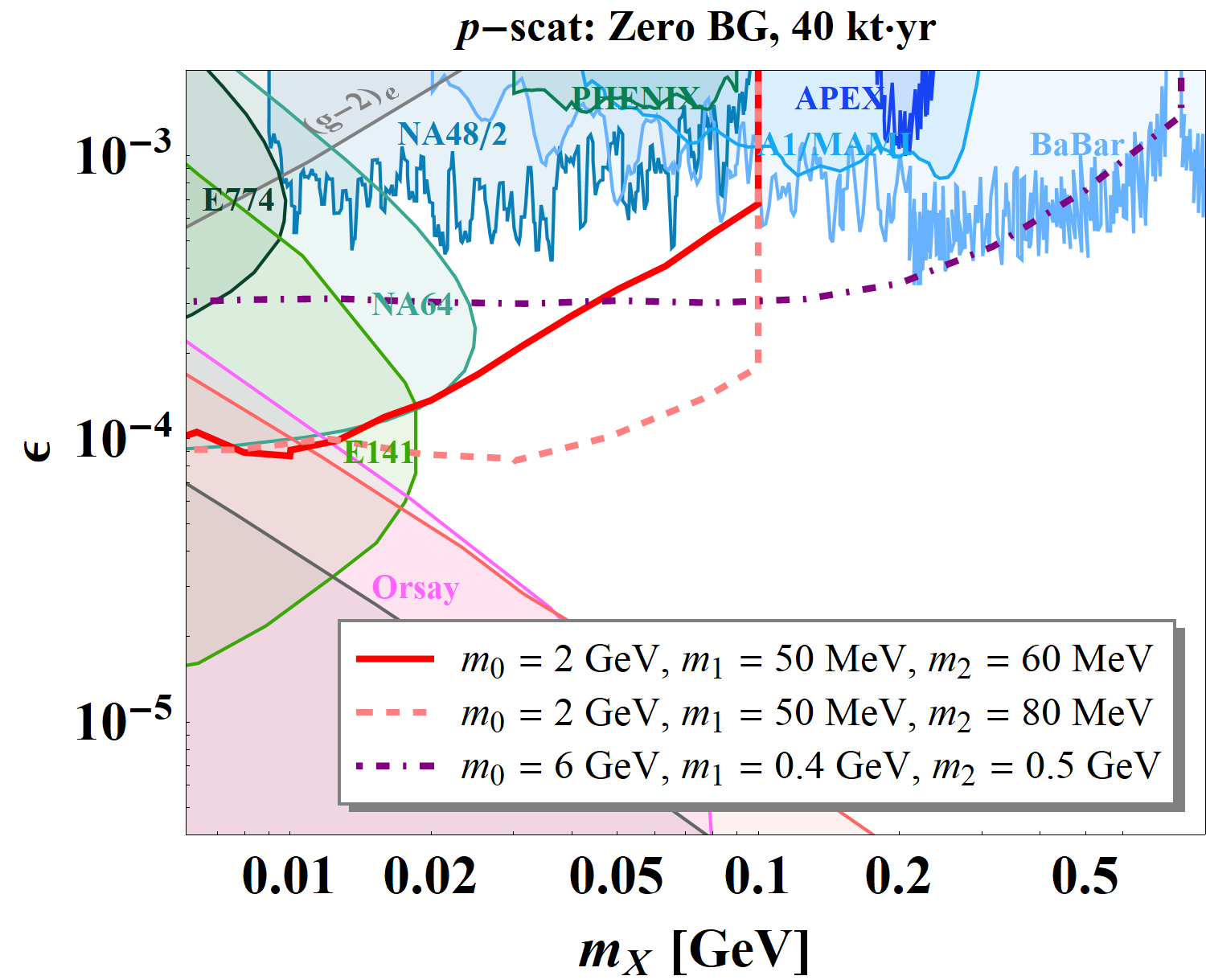} \hspace{0.2cm}
    \includegraphics[width=7.3cm]{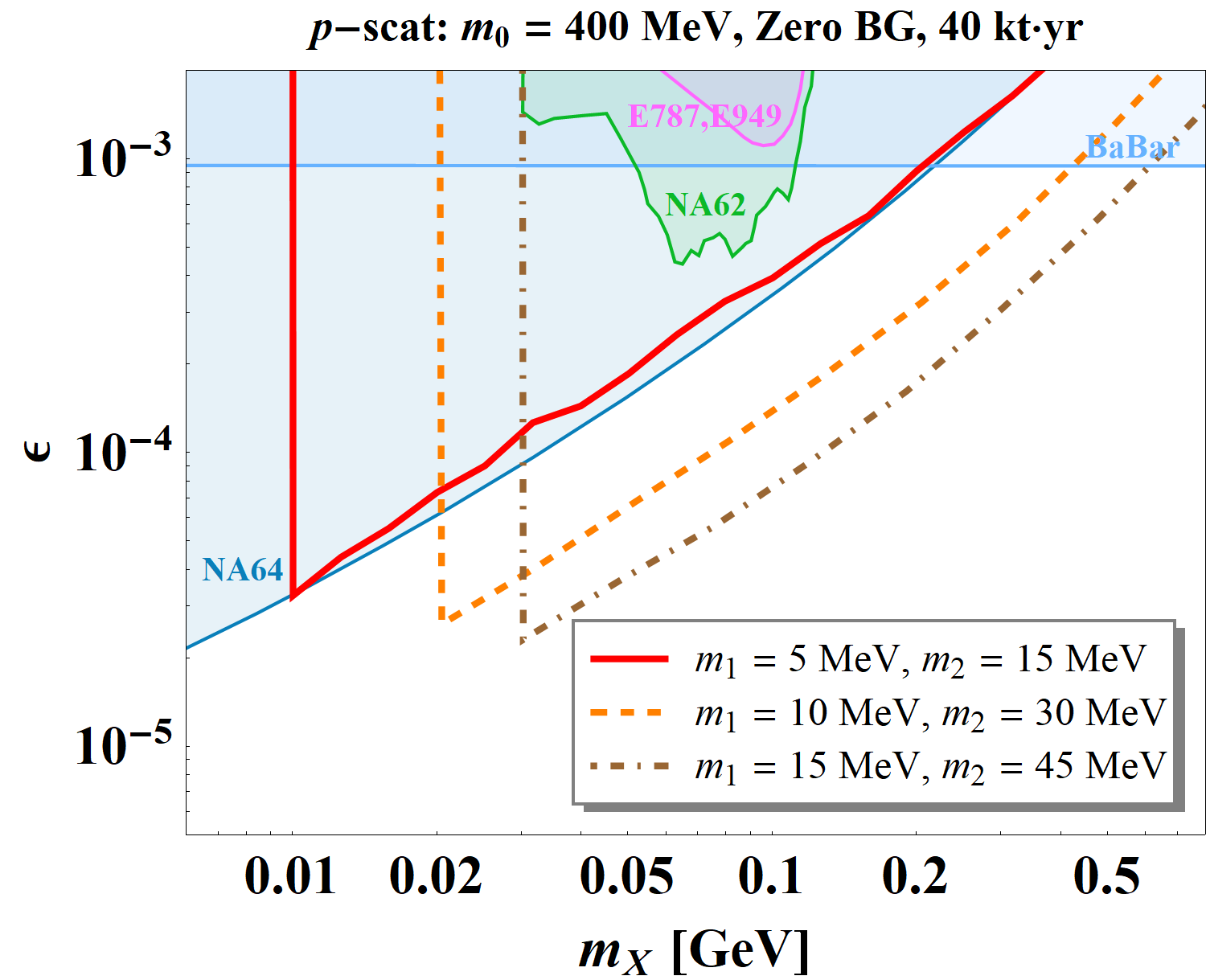}
    \caption{Left: Comparison between the sensitivity of DUNE to REF1 (with $m_X$ varied) and to two other model points in the proton scattering channel for the case of $m_X<2m_1$.
    Right: Comparison between the sensitivity of DUNE to REF2 (with $m_X$ varied) and to two other model points in the proton scattering channel for the case of $m_X \geq 2m_1$.
    In both panels, negligible background and 40-kt$\cdot$yr data collection are assumed, and parameter values and line specifications of the additional model points are detailed in the legends.}
    \label{fig:comp-pscat}
\end{figure}

Figure~\ref{fig:epsvsmX} shows that the proton scattering channel is more sensitive than the electron scattering channel if the underlying dark photon is heavier than a few tens of MeV, for which the kinetic mixing parameter is relatively loosely constrained. 
Furthermore, following eq.~\eqref{eq:m2prot}, the proton target offers a wider range of accessible $m_2$ values for a given pair of $m_1$ and $E_1$, allowing to carry out sensitivity studies for a larger range of parameter space.
As an illustration we analyze two more benchmark points for both the $m_X < 2m_1$ and $m_X \geq 2m_1$ case, and show the comparisons in the left panel and the right panel of figure~\ref{fig:comp-pscat}, respectively. 
The benchmark details are given in the legend of each figure, and the sensitivity reaches are computed under the assumption of negligible background and for a 40-kt$\cdot$yr exposure. 

The model point represented by the dashed orange line in the left panel differs from REF1 by $m_2$ value, showing that DUNE is sensitive to a broader range of $m_X-\epsilon$ space for this point, compared to the REF1 case (solid red).  
The reason is two-fold.
First, up to $m_X<m_2-m_1=30$~MeV, the whole signal process proceeds rather promptly, so it is highly probable that all three final-state particle tracks are fully contained.
Second, beyond $m_X=30$~MeV, $\chi_2$ decays through a virtual dark photon, hence becomes long-lived.
However, both $m_2$ and $m_2-m_1$ values are larger than those of REF1, resulting in a higher chance of containment of the $\chi_2$ decay vertex within the detector fiducial volume.
The dot-dashed purple line shows the sensitivity to a heavier mass spectrum.
In the right panel where $m_X \geq 2m_1$, we keep $m_0 (=E_1)$ fixed (i.e., the $\chi_1$ flux is fixed) but vary $m_1$ for a constant value of $(m_2-m_1)/m_2$.
In all cases, only the three-body decay of the $\chi_2$ is available, so larger mass gaps and larger $m_2$ values allow more events to be contained within the detector fiducial volume.
In addition, for $m_1 \ll m_p$, the energy transfer to the target proton is more efficient for larger value of  $m_1$~\cite{Kim:2020ipj}.
Therefore, for a given $m_X$, DUNE will be sensitive to smaller $\epsilon$ values in the two model points with larger $m_1$, extending into  unconstrained new regions of parameter space.
We emphasize that all these additional model points, other than REF1 and REF2, cannot be accessed in the electron scattering channel as they are kinematically forbidden due to larger mass gaps.
These studies illustrate that if it were possible in future to reduce the energy threshold for detecting protons, it would open further a powerful window to explore more exciting dark-sector scenarios with multiple (unstable) dark-sector states. 
For example, our simulation studies with the proton energy threshold being 21~MeV show that the sensitivity reaches in $\epsilon$ can be improved by $\sim 10-30$\%. 

\begin{figure}[t]
    \centering
    \includegraphics[width=7.3cm]{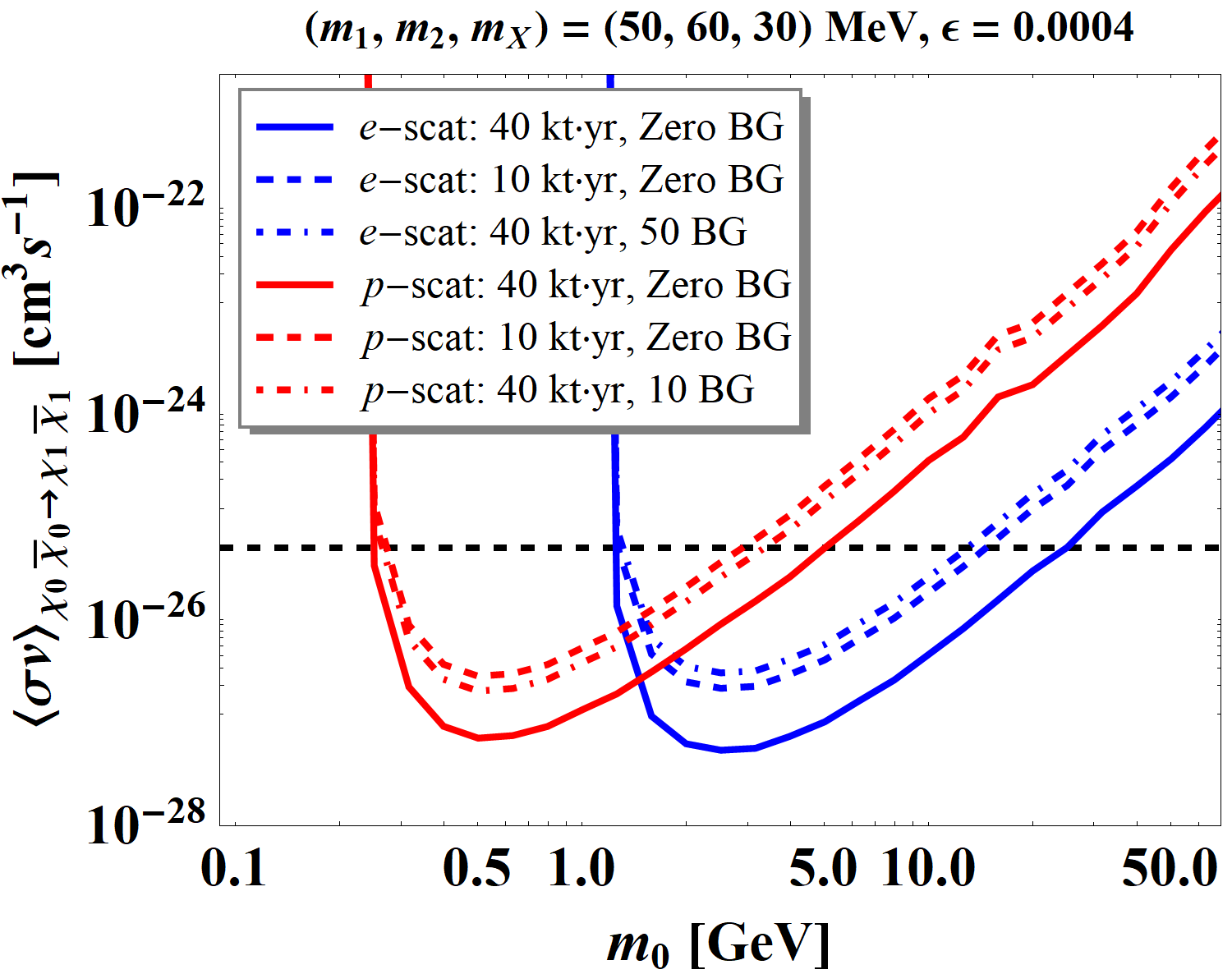} \hspace{0.2cm}
    \includegraphics[width=7.3cm]{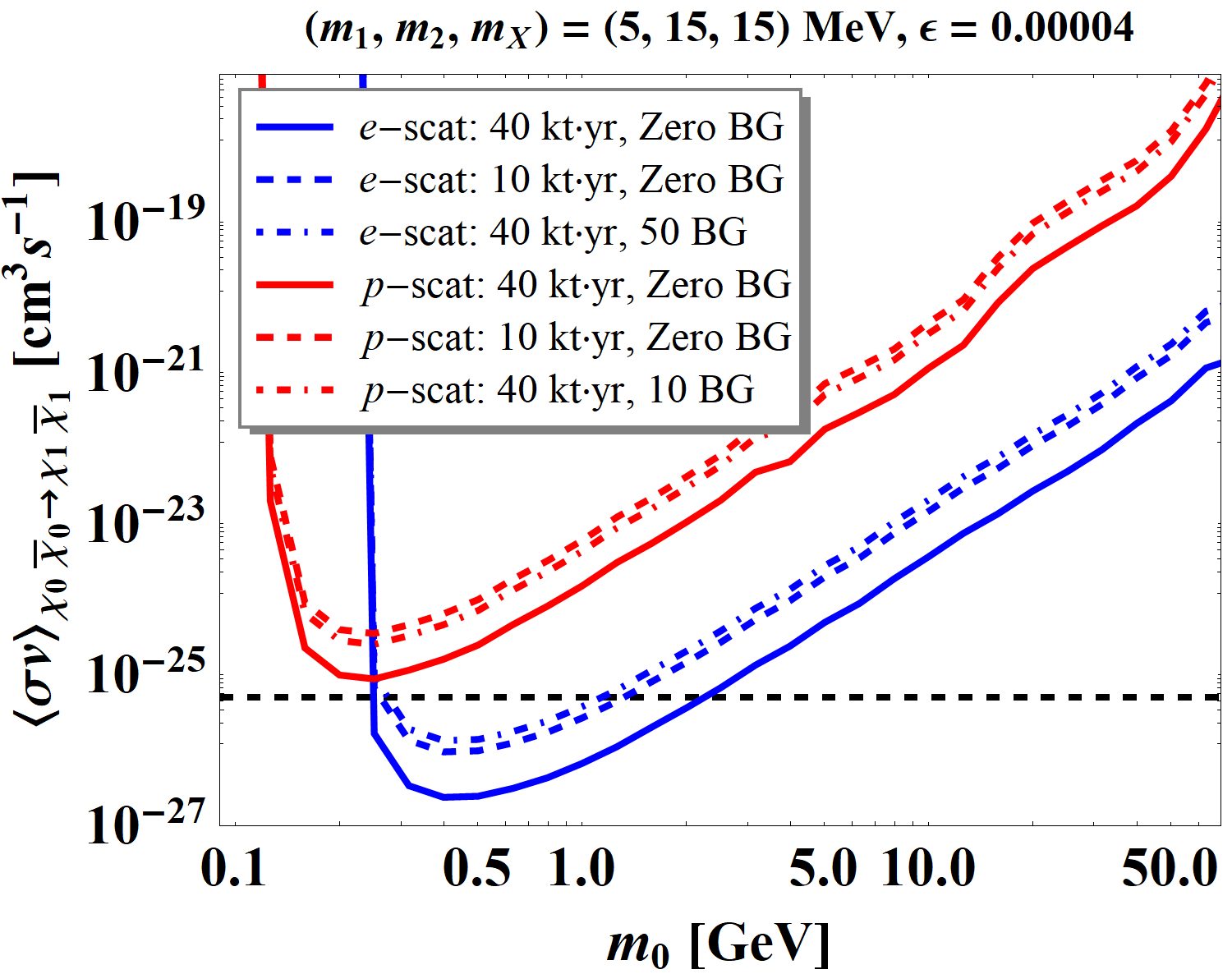}
    \caption{The expected sensitivity reaches of the velocity-averaged annihilation cross section for the $\chi_0\bar{\chi}_0\to\chi_1\bar{\chi}_1$ process as a function of halo dark-matter mass $m_0$.
    The used mass spectra correspond to REF1 (left panel) and REF2 (right panel) except that $m_0$ is varied and the $m_X-\epsilon$ pair is fixed as shown in the upper frames.
    The dashed black lines mark a reference value of the velocity-averaged annihilation cross section with which the conventional two-component BDM scenario generates the observed relic abundance.}
    \label{fig:m0vssigv}
\end{figure}

Next we turn our attention to a different study where the sensitivity reach for the velocity-averaged annihilation cross section for $\chi_0\bar{\chi}_0\to\chi_1\bar{\chi}_1$ is investigated as a function of $m_0$, the mass of the halo dark-matter component.
Using eqs.~\eqref{eq:fluxform} and \eqref{eq:Nsignal} together with $N^{90}$, we derive the sensitivity bound as follows:
\begin{equation}
    \frac{\langle \sigma v \rangle _{\chi_0\bar{\chi}_0 \rightarrow \chi_1\bar{\chi}_1}}{5\times 10^{-26}~{\rm cm}^3{\rm s}^{-1}} \geq \frac{N^{90}}{1.6\times10^{-6}{\rm cm}^{-2}{\rm s}^{-1}\left(\frac{10~{\rm GeV}}{m_0} \right)^2\ \sigma\ A_{\rm exp}\ t_{\rm exp} \ N_T}\,,
\end{equation}
for which a few examples are reported in figure~\ref{fig:m0vssigv}, in the plane of $m_0$ and $\langle \sigma v \rangle _{\chi_0\bar{\chi}_0 \rightarrow \chi_1\bar{\chi}_1}$.
This parameter space is reminiscent of presenting results from dark-matter indirect searches: the direct detection of a boosted $\chi_1$ scattering signal can be interpreted as an indirect detection of $\chi_0$ via its annihilation products.
The reference mass spectra are exactly the same as for REF1 (left panel) and for REF2 (right panel) except that $m_0$ is varied and the choices for $m_X$ and $\epsilon$ are not excluded by current bounds mentioned for figures~\ref{fig:epsvsmX} and \ref{fig:comp-pscat}.
The same scenarios as for figure~\ref{fig:epsvsmX} are considered, as explained in the figure legend.

A couple of remarks are in order. 
First of all, the proton channel, in general, allows to access smaller $m_0$ values than the electron channel because the proton target is better for $\chi_2$ production, with a smaller $E_1(= m_0)$, as discussed in section~\ref{sec:kin}.
However, if $m_0$ is too small, the energy deposited by recoiling protons is below threshold so that the sensitivity gets quickly degraded even though $m_2$ is kinematically allowed. 
Second, the dashed black lines mark the reference value of the velocity-averaged annihilation cross section, $5\times 10^{-26}~{\rm cm}^3{\rm s}^{-1}$, with which the conventional two-component BDM scenario reproduces the correct relic abundance~\cite{Belanger:2011ww}.
The model points below the line would lead to over-production of dark matter.
These results show that DUNE should be able to probe the dark-matter over-production limit in the context of the annihilating BDM scenario.  
A signal discovery below the limit would require a modification of cosmology in the early universe to accommodate the conventional BDM scenario. 

\subsection{Model-independent sensitivity reaches \label{sec:modelind}}

Returning to eq.~\eqref{eq:Nsignal}, we note that the model details and the dark-matter halo profile are encapsulated in $\sigma$ and $\mathcal{F}_1$, respectively. The acceptance $A_{\rm exp}$  depends on the underlying model details as well.
By contrast, the other two quantities $t_{\rm exp}$ and $N_T$ describe pure experimental conditions.
Reference \cite{Giudice:2017zke} suggested a possible way of presenting the experimental sensitivity in a model-independent manner.
We follow this suggestion here, based on the following idea. 
Since many well-motivated model points involve displaced vertices, the acceptance associated with the laboratory-frame decay length (of either $\chi_2$ or $X$) may be factored out like $A_{\rm exp}\equiv A(\ell_{\rm lab}) \tilde{A}_{\rm exp}$.
However, $\ell_{\rm lab}$ differs from event to event, hence a pragmatic theoretical choice is to take maximum laboratory-frame mean decay length $\bar{\ell}_{\rm lab}^{\max}$. 
The sensitivity reach is then formally expressed as
\begin{equation}
    \sigma_{\rm fid}\ \mathcal{F}_1 \geq \frac{N^{90}}{A(\bar{\ell}_{\rm lab}^{\max})\ t_{\rm exp} \ N_T}\,,
\end{equation}
where $\sigma_{\rm fid}$ denotes the ``fiducal'' cross section defined by $\sigma_{\rm fid}=\sigma  \tilde{A}_{\rm exp}$.
We estimate $A(\bar{\ell}_{\rm lab}^{\max})$ by requiring both the primary scattering vertex and the secondary decay vertex to be detectable in the detector fiducial volume, assuming an isotropic dark-matter signal flux. 
This sets a conservative limit since the laboratory-frame mean decay length of each event i.e., $\bar{\ell}_{\rm lab}^i$, is smaller than $\bar{\ell}_{\rm lab}^{\max}$ and, in turn, $A(\bar{\ell}_{\rm lab}^{\max})\leq \sum_i^{N_{\rm sig}} A(\bar{\ell}_{\rm lab}^i)/N_{\rm sig}$.

\begin{figure}[t]
    \centering
    \includegraphics[width=7.3cm]{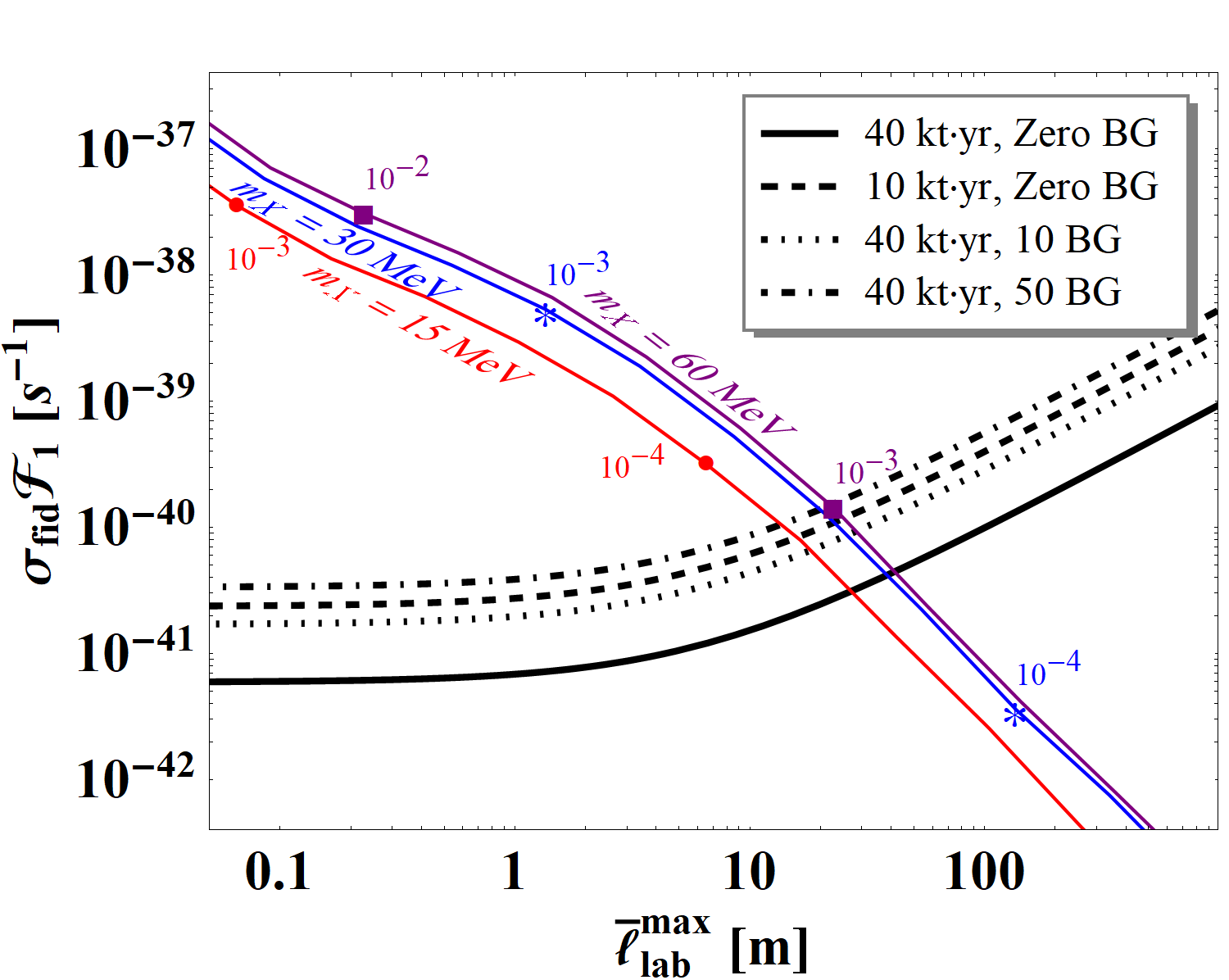} \hspace{0.2cm}
    \includegraphics[width=7.3cm]{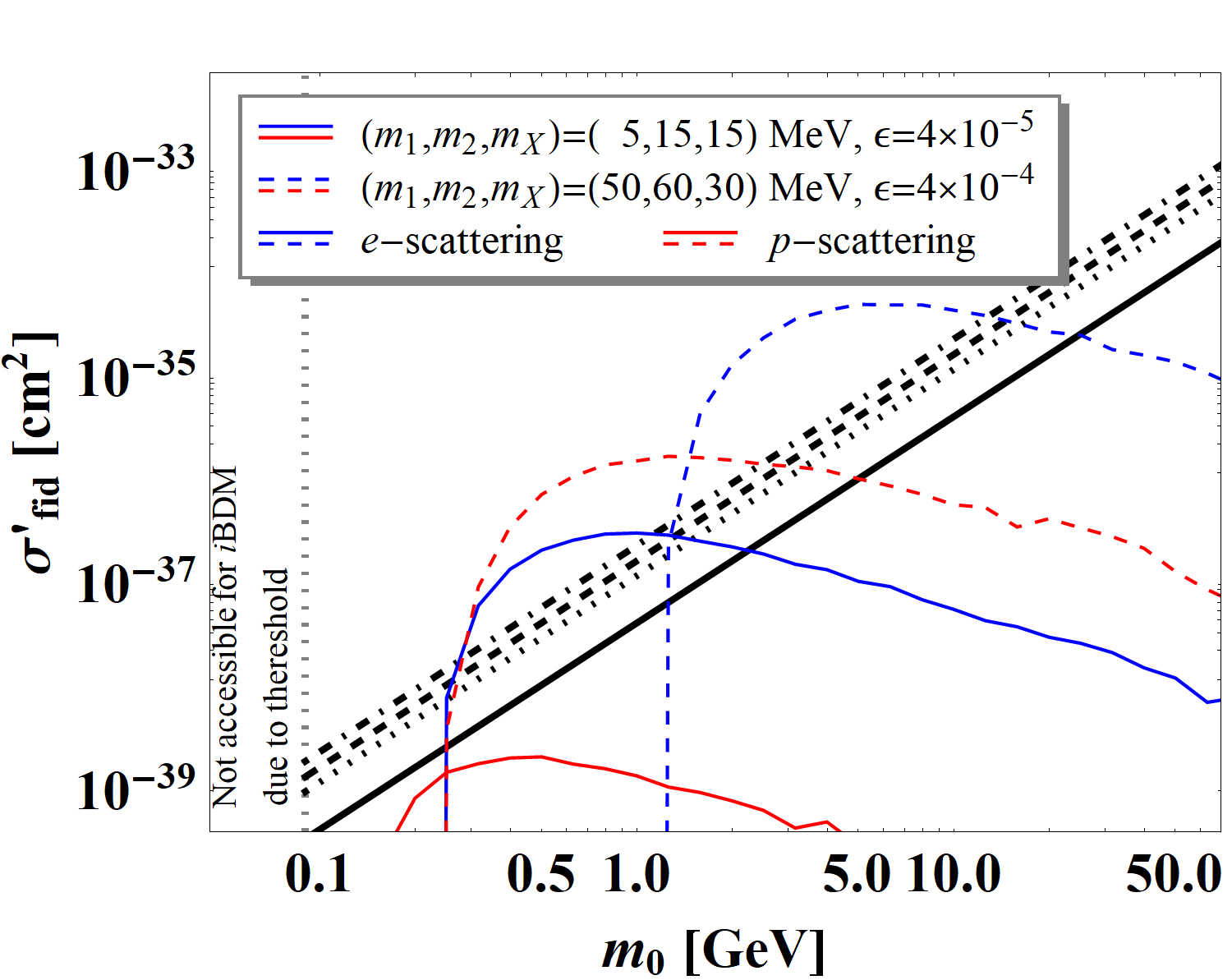}
    \caption{Left: Model-independent sensitivity reaches in the $\bar{\ell}_{\rm lab}^{\max}-\sigma_{\rm fid}\mathcal{F}_1$ plane for DUNE.
    The colored lines correspond to trajectories of $\sigma_{\rm fid}\mathcal{F}_1$ of REF1 in the electron channel.
    $\epsilon$ values are varied for three different choices of $m_X$ as shown in the plot.
    A few representative $\epsilon$ values are also displayed allowing to construct corresponding curves with a fixed $\epsilon$ and varying $m_X$ values by joining the point of each curve with the same $\epsilon$.
    Right: Model-independent sensitivity reaches in the $m_0 -\sigma'_{\rm fid}$ plane for DUNE with $\sigma'_{\rm fid}$ identified as $\sigma_{\rm fid}A(\bar{\ell}_{\rm lab}^{\max})$. 
    The styles of the black lines are the same as in the left panel. 
    The colored lines depict the expected $\sigma'_{\rm fid}$ values in $m_0$ for the reference points adopted in figure~\ref{fig:m0vssigv} as also specified in the legend.}
    \label{fig:indlimits}
\end{figure}

The expected model-independent sensitivity reach of DUNE is displayed in the left panel of figure~\ref{fig:indlimits}: 40~kt$\cdot$yr-Zero BG (solid black), 10~kt$\cdot$yr-Zero BG (dashed black), 40~kt$\cdot$yr-10 BG (dotted black), and 40~kt$\cdot$yr-50 BG (dot-dashed black).
Note that these results are applicable to both the electron and the proton channels as $A(\bar{\ell}_{\rm lab}^{\max})$ is evaluated irrespective of the channel choice.
For a given model point, one can calculate a fiducial cross section multiplied by the expected signal flux to check whether or not it is excluded.
For illustration, we calculate products of fiducial cross sections for REF1 in the electron channel and the signal flux predicted with the NFW dark-matter halo profile, while varying $\epsilon$ for three representative $m_X$ choices, $m_X=15$~MeV (red), $m_X=30$~MeV (blue), and $m_X=60$~MeV (purple).
A few reference $\epsilon$ values are also shown.
The model points along the line segment above (below) a given exclusion curve are ruled out (allowed).
Similar curves with a fixed $\epsilon$ and varying $m_X$ can be obtained by joining the point of each curve with the same $\epsilon$ value: for example, a line connecting the points of $\epsilon=10^{-3}$ in the plot.

While this presentation scheme is interesting {\it per se}, there is another way to report the experimental sensitivity in a more familiar fashion by reintroducing the dependence of the dark-matter halo distribution encoded in $\mathcal{F}_1$:
\begin{equation}
    \sigma'_{\rm fid} \geq \frac{N^{90}}{\mathcal{F}_1(m_0)\ t_{\rm exp}\ N_T}\,,
\end{equation}
where $\sigma'_{\rm fid}=\sigma_{\rm fid}A(\bar{\ell}_{\rm lab}^{\max})$.
Here we explicitly indicate the dependence of the signal flux on the mass of relic dark matter $\chi_0$.
The resulting sensitivity is  defined in the $m_0-\sigma'_{\rm fid}$ plane which is reminiscent of the limits of spin-independent and spin-dependent cross sections as a function of the mass of the dominant relic dark matter particle in conventional dark matter direct detection experiments.

The expected sensitivity is shown  in this presentation in the right panel of figure~\ref{fig:indlimits} and the line styles are identical to those in the left-panel plot.
As before, the NFW dark-matter profile is applied and $\langle \sigma v \rangle _{\chi_0\bar{\chi}_0 \rightarrow \chi_1\bar{\chi}_1}$ is set to be $5\times 10^{-26}~{\rm cm}^3{\rm s}^{-1}$.
The black vertical dotted line marks the absolute lower bound for visible triple track events due to the energy threshold.
Again these results are applicable to both the electron and the proton channels since no channel details are assumed.
Similarly, one can check whether or not a given model point is ruled out by calculating the fiducial cross section associated with $\sigma'_{\rm fid}$.
Example fiducial cross sections for kinematically consistent $m_0$ values are shown in the plot.
The chosen mass spectra and $\epsilon$ values are the same as in figure~\ref{fig:m0vssigv} as indicated in the legend.
Basically, line segments above (below) a given black diagonal line may be ruled out (allowed) by DUNE.
All example points have a range of $m_0$ values that can be ruled out except the benchmark point represented by the solid red line.
This should be compared with the proton scattering case in the right panel of figure~\ref{fig:m0vssigv} which does not reach the line of $\langle \sigma v \rangle _{\chi_0\bar{\chi}_0 \rightarrow \chi_1\bar{\chi}_1}=5\times 10^{-26}~{\rm cm}^3{\rm s}^{-1}$.

\section{Conclusions \label{sec:conclusion}}

Dark matter and neutrino oscillations are evident signs of physics beyond the Standard Model.
To study the mysteries of neutrinos, many neutrino experiments are ongoing and several are being planned for the near future.
In particular, large-volume neutrino experiments such as DUNE~\cite{Abi:2020wmh,Abi:2020evt} and HK~\cite{Abe:2016ero,Abe:2018uyc} are expected to take
the lead towards new groundbreaking observations and discoveries in the next 10 years.

Due to the common challenge of the invisible and feebly interacting nature that dark matter and neutrinos share, one can opportunistically anticipate that these neutrino experiments have excellent capabilities of detecting certain classes of dark-matter signals.
The large-volume detectors exhibit particle energy thresholds in the tens of MeV range, hence they do not have a significant sensitivity to conventional non-relativistic dark matter but rather to experimental signatures induced by, for example, relativistic dark matter.
An increasing number of non-conventional dark-matter scenarios or models have been proposed during the last years~\cite{DEramo:2010keq,Belanger:2011ww,Huang:2013xfa,Agashe:2014yua,Berger:2014sqa,Kong:2014mia,Kim:2016zjx,Kim:2017qaw,Aoki:2018gjf,Bringmann:2018cvk,Ema:2018bih,Dent:2019krz,Bhattacharya:2014yha,Kopp:2015bfa,Heurtier:2019rkz} and they postulate the presence of relativistic light dark matter in the universe at the present time. 
Due to its relativistic nature, such cosmogenic dark matter can manifest itself in the detectors as an energetic visible target recoil, accompanied by additional visible particles, depending on the underlying dark-sector model details.

In this paper, we have studied the sensitivity of a detector similar to the one proposed by the DUNE Collaboration to dark-matter signatures that involve multiple particle production in the final state, taking an inelastic boosted dark-matter scenario~\cite{Kim:2016zjx} as a concrete example.
In this scenario, the underlying dark sector minimally consists of a heavy dark matter $\chi_0$, a light dark matter $\chi_1$, an unstable dark-sector state $\chi_2$ (heavier than $\chi_1$), and a dark photon $X$ mediating the interactions among $\chi_1$, $\chi_2$, and SM particles.
An incident $\chi_1$, which is boosted by pair-annihilation of the dominant and much heavier relic dark matter $\chi_0$ in the galaxy, scatters off an electron or a proton in the DUNE far detector volume to produce a $\chi_2$.
The collision produces a recoiling electron or a recoiling proton together with additional SM particles, and as in this study an electron-positron pair coming from the decay of $\chi_2$ through an intermediary state $X$.
The presence of additional particles gives several unique event signatures
which can be used to distinguish signal from background.
But for that the detector should have a good particle isolation/identification and exhibit excellent energy and angular resolutions. 
In this sense, the DUNE far detectors based on the LArTPC technology can meet these requirements so that they can obtain highly competitive experimental 
sensitivities to these dark-matter scenarios.

We first studied the energy spectra and angular separation of final-state particles for two representative reference points in the inelastic boosted dark-matter scenario.
Recoiling electrons and secondary electrons/positrons (i.e., $e^\pm$ from the $\chi_2$ decay) are typically energetic enough to pass the detector energy threshold for electrons.
In contrast, light dark matter $\chi_1$ interacting with protons typically invokes a small energy transfer to the recoiling proton, requiring a small kinetic-energy threshold for protons.
On the other hand, the angular spectra demonstrate that the final-state particles are likely to get merged and collimated in the electron scattering case, while they are rather separated and isolated in the proton scattering case.
We showed that a $dE/dx$-based analysis can help to recognize 
merged tracks, and performed the first study on the expected $dE/dx$ distributions for multi-particle merged tracks.

We then performed a sensitivity study for the benchmark model, simulating signal events at several model points.
Potential background sources were identified and estimated, based on the expected performance of the DUNE LArTPC far detectors.
We have also defined a selection scheme for the detector-level signal events.
Various detector effects such as energy thresholds, resolutions, smearing, and particle track lengths were parameterized. 

The sensitivity reach for the conventional dark-photon parameter space was first investigated and it was also compared for two model points with the existent $i$BDM sensitivity reach reported by the COSINE-100 Collaboration.
Our study showed that the DUNE far detectors have an excellent potential to probe unexplored regions of dark-matter parameter space.
In particular, searches in the proton channel are very promising in terms of exploring a wide range of non-minimal dark-sector scenarios.
This is encouraging, and suggests to aim for further improvements in the proton identification of the DUNE LArTPC detectors, for lower kinetic energies.
We have also studied the sensitivity reach in the velocity-averaged annihilation cross section for halo dark matter $\chi_0$ for a given mass value of the $\chi_0$.
Our results show that DUNE would be able to reach sensitivity into the dark-matter over-production region, which can be set by the assisted freeze-out mechanism~\cite{Belanger:2011ww}, for the conventional two-component boosted dark-matter scenario. 

We have also presented the results of the experimental reaches in a model-independent manner.
Two presentation schemes were discussed: one in the $\bar{\ell}_{\rm lab}^{\max}-\sigma_{\rm fid}\mathcal{F}_1$ plane and the other in the $m_0-\sigma'_{\rm fid}$ plane.
The former is motivated for typical signal events accompanied by a displaced vertex signature, while the latter is analogous to the presentations of limits of the spin-(in)dependent cross sections as a function of the halo dark-matter mass in conventional dark matter direct detection experiments.
For a given model point, one can check whether or not it is excluded by these limits.

Finally, we emphasize that our study here can be readily generalized to generic signal events containing a large multiplicity of final-state particles, not just limited to the benchmark dark-sector scenario that we have considered.
We encourage the DUNE experiment to pioneer exploring non-minimal dark-sector scenario searches, and contribute in a major way to shed light on dark-matter physics, presently one of the key science questions in fundamental physics and cosmology.

\section*{Acknowledgments}
We thank Soo-Bong Kim for insightful/useful discussions.
The work of DK was supported in part by the Department of Energy under Grant DE-FG02-13ER41976 (de-sc0009913) and is supported in part by the Department of Energy under Grant de-sc0010813.
The work of JCP is supported by the National Research Foundation of Korea (NRF-2019R1C1C1005073 and NRF-2018R1A4A1025334). 
The work of SS was supported by the National Research Foundation of Korea (NRF-2020R1I1A3072747). 
This work was performed at the Aspen Center for Physics, which is supported by National Science Foundation grant PHY-1607611.
SS would like to express a special thanks to the Mainz Institute for Theoretical Physics (MITP) of the Cluster of Excellence PRISMA+ (Project ID 39083149) for its hospitality and support.

\appendix

\section{Decay width of $\chi_2$ \label{sec:app}}

We provide the exact formula of $\Gamma_2$ for the case of $\chi_2 \rightarrow \chi_1 X^* \rightarrow \chi_1 e^+e^-$. We refer to ref.~\cite{Giudice:2017zke} for the detailed derivation. 
\bea
\Gamma_2 = \frac{g_{12}^2\epsilon^2 \alpha}{64 \pi^2 m_2^3}\int_{s_2^-}^{s_2^+}ds_2\int_{s_1^-}^{s_1^+}ds_1 \frac{\overline{|\mathcal{A}|}^2}{\left(m_1^2+m_2^2+2m_e^2-s_1-s_2-m_X^2\right)^2+m_X^2\Gamma_X^2}\,,
\eea
where $\overline{|\mathcal{A}|}^2$ in our benchmark model~\eqref{eq:laglangian} is given by
\bea
\overline{|\mathcal{A}|}^2&=&4\left\{(s_1+s_2)\left[(m_1+m_2)^2+4m_e^2 \right]-(s_1^2+s_2^2)-2m_1 m_2(m_1^2+m_2^2+m_1m_2) \right. \nonumber \\
&-& \left. 2m_e^2(m_1^2+m_2^2+4m_1m_2+3m_e^2) \right\}\,.
\eea
Here the integration limits are 
\bea
s_1^\pm &=& m_1^2+m_e^2+\frac{1}{2s_2}\left[ (m_2^2-m_e^2-s_2)(m_1^2-m_e^2+s_2)\pm \lambda(s_2, m_2^2,m_e^2)\lambda(s_2, m_2^2, m_e^2)\right], \nonumber \\
s_2^- &=& (m_1+m_e)^2\,, \hbox{ and }s_2^+=(m_2-m_e)^2\,,
\eea
with $\lambda(x,y,z)\equiv \sqrt{x^2+y^2+z^2-2(xy+yz+zx)}$.



\bibliographystyle{JHEP}
\bibliography{ref}

\end{document}